\documentclass{rspublic}
\usepackage{amsmath}
\usepackage{amsfonts}
\usepackage{amssymb}
\usepackage[Symbol]{upgreek}
\usepackage[nointegrals]{wasysym}
\usepackage{mathrsfs}
\usepackage{graphicx}
\newcommand{\bfu}{{\bf u}}
\newcommand{\bfx}{{\bf x}}
\begin{document}
\title[One Dimensional Navier-Stokes Model]{\large Non-linear vorticity upsurge in Burgers' flow} 
\author[F. Lam]{F. Lam}
\label{firstpage}
\maketitle
\begin{abstract}{Navier-Stokes Regularity; Burgers equation; Viscosity; Vorticity; Laminar-turbulent Transition; Non-linearity; Shock; Turbulence Decay}
We demonstrate that numerical solutions of Burgers' equation can be obtained by a scale-totality algorithm for fluids of small viscosity (down to one billionth). Two sets of initial data, modelling simple shears and wall boundary layers, are chosen for our computations. Most of the solutions are carried out well into the fully turbulent regime over finely-resolved scales in space and in time. It is found that an abrupt spatio-temporal concentration in shear constitutes an essential part during the flow evolution. The vorticity surge has been instigated by the non-linearity complying with instantaneous enstrophy production, while {\it ad hoc} disturbances play no role in the process. In particular, the present method {\it predicts} the precipitous vorticity re-distribution and accumulation, predominantly over localised regions of minute dimension. The growth rate depends on viscosity and is a strong function of initial data. Nevertheless, the long-time energy decay is history-independent and is inversely proportional to time. Our results provide direct evidence of the vorticity proliferation embedded in the equations of motion. The non-linear intensification is a robust feature, and is ultimately responsible for the drastic succession in boundary layer profiles over the intrinsic laminar-turbulent transition (Schubauer \& Klebanoff 1955). The dynamical inception of turbulence can be decrypted by solving the full time-dependent Navier-Stokes equations which ascribe no instability stages.
\end{abstract}
%
%
%
%
\section{Introduction}
It has been long known that fluid motions start from a streamlined laminar state, undergo {\it laminar-turbulent transition}, and evolve into randomly fluctuating poly-scaled turbulence (see the experiments of Reynolds 1883; Taylor 1923; Schubauer \& Skramstad 1947). In the Eulerian description of the motion of an incompressible, homogeneous Newtonian fluid, the Navier-Stokes equations of motion and the continuity read
\begin{equation} \label{ns}
	{\partial {\bfu}}/{\partial t} + ({\bfu} . \nabla) {\bfu} = \nu \Delta {\bfu}   - {\rho}^{-1} \nabla p, \;\;\; \nabla . {\bfu} =0,
\end{equation}
where vector ${\bfu}={\bfu}({\bfx},t)$ is the velocity having the components $(u_1,u_2,u_3)$, the scalar quantity $p=p({\bfx},t)$ denotes the pressure, ${\bfx}=(x_1,x_2,x_3)$ the space variable, and $\Delta$ the Laplacian. The kinematic viscosity is $\nu =\mu/\rho$, where $\rho$ and $\mu$ are the density and viscosity respectively. 
We are interested in the initial-boundary value problem of (\ref{ns}) from given data of finite energy
\begin{equation} \label{ic}
 {\bfu}({\bfx},t{=}0) = {\bfu}_0({\bfx}) \; \in \; C_c^{\infty}(\Upomega),
\end{equation} 
and $\nabla.\bfu_0=0$. The no-slip boundary condition applies on the (smooth) boundary of the domain $\Upomega$
\begin{equation} \label{bc}
 {\bfu}({\bfx},t) = 0\;\;\; \forall \bfx \in \partial \Upomega.
\end{equation} 
Taking divergence of (\ref{ns}) and utilising the continuity, we derive a Poisson equation for the pressure 
\begin{equation} \label{pp}
	\Delta p(\bfx)= - \rho \sum^{3}_{i,j=1} \Big( \frac{\partial u_j}{\partial x_i} \: \frac{\partial u_i}{\partial x_j}\Big)(\bfx)
\end{equation}
the solution of which is well-known. A physical vector quantity, $\omega{=}\nabla {\times} \bfu$, is the vorticity which is related to the angular momentum in flow motions. The vorticity equation is
\begin{equation} \label{vort}
	{\partial \omega}/{\partial t} - \nu \Delta \omega = (\omega . \nabla) \bfu  - (\bfu . \nabla )\omega.
\end{equation}
The vorticity field inherits the solenoidal property of (\ref{ns}) $\nabla. \omega=0$. At every given instant of time, we recover the velocity from the di-vorticity from the Biot-Savart relation
\begin{equation} \label{bs}
\bfu(\bfx)= \Delta^{-1} (\nabla \times \omega),
\end{equation}
where $\Delta^{-1}$ stands for the Laplacian inverse in $\Upomega$ satisfying (\ref{bc}). This is an elliptic relation, just like the solution of (\ref{pp}). Neither contains time information. 

To solve dynamic problems for fluid motions, we may tackle either (\ref{ns})-(\ref{pp}) or (\ref{vort})-(\ref{bs}) with appropriate initial and boundary data. In theory, each set of the equations defines a parabolic-elliptic system by the assumption of incompressibility. The Navier-Stokes equations describe the mean velocity as well as the pressure (in the sense of ensemble averages) and do not model the random fluctuations in turbulent motions. For practical purposes, it is precisely the motive underlying the continuum hypothesis where only mean flow quantities, such as skin friction and average energy dissipation increase over transition, really matter. 

The study of the transitional dynamics is to address the mechanism of {\it en masse} scale initiation by the term $(\bfu.\nabla) \bfu$. Consequently, the causality of external disturbances on flow development is of secondary importance; the effects of extraneous forcing can only be well-quantified once the intrinsic transition has been understood. In the present paper, an effort is made to analyse the non-linear dynamics by solving Burgers' equation numerically. We shall present our numerical evidence for the intrinsic process in small-viscosity flows. By examining the solutions, we are able to identify the hallmarks of the transition process as observed in experiments. 
\section{Burgers' equation and mathematics of turbulence}
Burgers' equation refers to the parabolic equation in one-space dimension (Burgers 1948):
\begin{equation} \label{beqn}
\frac{\partial u} {\partial t} - \nu \; \frac{\partial^2 u}{\partial x^2} = - u \frac{\partial u}{\partial x}.	
\end{equation}
We seek its solutions subject to initial-boundary conditions,
\begin{equation} \label{bic}
	u(x,t=0)=u_0(x), \;\;\; \mbox{and} \;\;\; u(0,t)=u(1,t)=0.
\end{equation}
Two initial data examined in the present paper are given by
\begin{equation} \label{bl}
	u_0(x)=A_0\sin(\pi x),
\end{equation}
which evolves into a boundary layer near $x=1$ for constant intensity $A_0>0$, and
\begin{equation} \label{shear}
	u_0(x)=A_0\sin(2 \: \pi x),
\end{equation}
which has the time-wise development resembling the formation of a shear layer or a standing saw-tooth shock in the vicinity of $x=0.5$. 

In many textbooks on differential equations, the following example is often cited to illustrate the existence of blow-up solutions. Let u=u(t) be governed by the ordinary differential equation 
\begin{equation*} 
	{\rd u }/{\rd t} = u^2,\;\;\;\;\; u(t=0)=u_0.
\end{equation*} 
By means of substitution $u=1/v$, the solution is found to be $u(t)= {u_0}/{(1 - u_0 t)}$ which blows up at $t=1/u_0$ for positive initial data. For finite $t<1/u_0$, the solution is regular. Taking $u$ as velocity, the blow-up scenario contradicts the laws of thermodynamics. We may blame the negligence of certain dissipation in our model. In part, the contradiction is due to our mathematical analysis. Consider the motion in time reversal (or the adjoint in a sense), $t \rightarrow -t$. Now the blow-up solution becomes $u(t)= {u_0}/{(1 + u_0 t)}$ which is globally regular for the identical positive data. The first law suggests that the blow-up option must be discarded, in the same way as choosing exponentially damped eigen-states. Similar principles hold for equations containing a cubic non-linearity ${\pm}u^3$. The singularity scenarios violating the thermodynamic principles have no justifications in classical physics. 

The quadratic non-linearity in (\ref{beqn}) epitomises a number of natural phenomena in continuum. Its analytical properties have been exploited and elucidated for prototype problems in turbulence, shock formation and aerodynamics (see, for instance, Burgers 1974; Whitham 1974). Roughly speaking, the non-linear advection-diffusion term has an identical form to those in the Navier-Stokes equations; it represents the energy transport per unit mass on the macroscopic scale:  
\begin{equation*}
	u \partial_x u = \partial_x \Big( \: \frac{1}{2}\:{u^2} \:\Big). 
\end{equation*}
Then the energy integral reads
\begin{equation} \label{intbd}
	\int_0^1 u \partial_x u \rd x = \frac{1}{2}\:{u^2} \:\Big|_0^1=0 
\end{equation}
in view of the boundary conditions. The flux is determined by the normal gradients on the boundary
\begin{equation} \label{flux}
	\frac{\rd}{\rd t }\int_0^1 u(x,t) \rd x = - {\nu}\; \frac{\partial u}{\partial x}\:\Big|_0^1,
\end{equation}
that is conserved only in flows satisfying $\partial_x u(0,t)=\partial_x u(1,t)$ (sometimes known as the second boundary value problems).
The shear stress on the `wall' $x=1$ per unit density for (\ref{bl}) is given by
\begin{equation} \label{wallshear}
	\tau_{w} = {\nu}\; \frac{\partial u}{\partial x}\:\Big|_{w}.
\end{equation}
Similarly, we are interested in the local shear at the `centre' of (\ref{shear}) ${\partial u}/{\partial x}\:|_{x=0.5}$.

Multiplying (\ref{beqn}) by $u$ and integrating the result over space, we obtain the energy conservation law
\begin{equation} \label{energy} 
	\frac{1}{2} \int_0^1 u^2 \; \rd x + \nu \int_0^t \Big( \int_0^1 \Big( \frac{\partial u}{\partial x} \Big)^2 \rd x \Big) \rd t  \; = \; \frac{1}{2} \int_0^1 u_0^2 \; \rd x.
\end{equation} 
In addition, the energy law (\ref{energy}) shows that the velocity at any instant must satisfy  
\begin{equation} \label{vbound}
	|u(x,t)| \leq A_0, \;\;\;\;\;\; 0 < x < 1,\;\;\; t > 0.
\end{equation}
This bound may serve as a useful criterion to monitor computations. For convenience, we shall call the quantity $\partial_x u/\partial x$ vorticity, as it appears in the enstrophy relation in (\ref{energy}) even though it may be a strain on face value.

Green's function for the heat operator in (\ref{beqn}) satisfying boundary conditions (\ref{bic}) is given by (for $t>0$)
\begin{equation} \label{grn}
  \begin{split}
	G(x,y,t) & = 2 \; \sum_{k=1}^{\infty} \; \sin(k \pi x) \; \sin(k \pi y) \; \exp\big( - \:\pi^2 k^2 \: \nu t\big)\\
	\quad & = \frac{1}{\sqrt{4 \pi \nu t}} \sum_{k=-\infty}^{\infty} \Big[ \exp \Big({-}\frac{(x{-}y{-}2k)^2}{4 \nu t}\;\Big) {-} \exp\Big( {-}\frac{(x{+}y{-}2k)^2}{4 \nu t} \;\Big) \Big].
	\end{split}
\end{equation}
By the summability of non-linearity (\ref{intbd}), we are justified to make use of Duhamel's principle. Burgers' equation then satisfies the following integral equation
\begin{equation} \label{bie}
	u(x,t) = w_0(x,t) + \int_0^t \int_0^1 K(x,y,t{-}s) \; u^2(y,s) \rd y \rd s,
\end{equation}
where
\begin{equation*}
	w_0(x,t)=\int_0^1 G(x,y,t) \; u_0(y) \rd y,
\end{equation*}
and the kernel function $K={\partial G}/{\partial y}/2$. As shown in \S 8 and \S12 of Lam (2013), the solution of (\ref{bie}) can be expressed in terms of a convergent series
\begin{equation} \label{ssol}
 \begin{split}
	u(x,t) = \; & \gamma^*(x,t) \:  + 2 \: V[\gamma^*]^2 + 10 \: V[\gamma^*]^3 + 62 \: V[\gamma^*]^4 + 430 \: V[\gamma^*]^5 + 3194 \: V[\gamma^*]^6 \\ 
	 \quad & \\
	\quad &  + 24850 \: V[\gamma^*]^7 + 199910 \: V[\gamma^*]^8 
	+ 1649350 \: V[\gamma^*]^9 + \: \cdots \cdots \: , 
 \end{split}
\end{equation}
where $\gamma^*(x,t)$ satisfies the linear Volterra-Fredholm integral equation
\begin{equation*} 
	\gamma^*(x,t) = w_0(x,t) + \int_0^t \int_0^1  K (x,y,t{-}s) \; \gamma^* (y,s) \rd y \rd s.
\end{equation*}
As a motion initiated from given finite data marches in time, more and more terms must be retained to represent the flow field. At every time $t>0$, an increasing amount of vorticity of smaller scales must be generated by the non-linearity ($u \partial_x u$) in order for the law of energy conservation to be fulfilled, thus producing enhanced local shear thanks to the presence of viscosity. Effectively, the flow evolves into a collection of vortices of various lengths and time scales. Every scale, 
$V[\gamma^*]^k $, has a precise meaning, and is defined by its space-time convolution of the initial data. This non-linear process of activating scales in abundance to balance instantaneous enstrophy production is described as {\it vorticity or scale proliferation}.
  
Evidently, it is a challenge to numerically evaluate all the terms of the series. In practice, one would prefer to bypass the direct evaluations. It is much easier to solve the dynamic equation (\ref{beqn}) by efficient algorithms. Recall that one will certainly avoid tedious calculations of the Fredholm analytic expression for integral equations at face value; it is definitely beneficial to solve the governing differential equation by straight-forward numerical techniques. 

The initial velocity (\ref{bl}) or (\ref{shear}) will evolve into space in time as $u=u(x,t)$. There does not exist a steady velocity throughout the motion by virtue of (\ref{energy}) and (\ref{vbound}). Lack of a characteristic velocity suggests that the flow development for $t>0$ cannot be expounded on dynamic similarity. On the basis of rigour, we do not identify viscosity as the unit Reynolds number $Re=1/\nu$ in the initial-boundary problem. This view asserts that any time-independent flow generated in laboratory or in nature, if exists, must be preceded by a time-dependent dynamics which galvanises contiguous shear diversity.
\section{Exact solutions at moderate viscosity}
General solutions of Burgers' equation can be written in a closed analytical form (Cole 1951; Hopf 1950).
Equation (\ref{beqn}) may be expressed in a conservative law 
\begin{equation} \label{beqn2}
	\frac{\partial u}{\partial t} = \frac{\partial }{\partial x }\Big(  \nu \; \partial_x u - \frac{1}{2} u^2 \Big).
\end{equation}
This formula suggests there exists a `similarity function' $\psi(x,t)$ so that
\begin{equation*} 
	u(x,t)=a \partial_x \psi(x,t)
\end{equation*}
for some constant $a$. A time differentiation shows the function is compatible to (\ref{beqn2}) or 
\begin{equation} \label{psix}
	\big( \partial_t - \nu \; \partial_{xx} \big) \psi = a \big( \partial_x \psi \big)^2 / 2.
\end{equation}
In order to eliminate the non-linear right-hand, we introduce $\psi = - b \log \phi (x,t)$. By differentiation, we can express $u$ in terms of $\phi$. Thus equation (\ref{beqn2}) is reduced to
\begin{equation*}
	-b \frac{\partial_t \phi}{\phi} = \nu b \Big( \frac{\partial_x \phi}{\phi}\Big)^2 - \nu b \frac{\partial_{xx} \phi}{\phi} - \frac{ab^2}{2}\Big( \frac{\partial_x \phi}{\phi}\Big)^2,
\end{equation*}
where it is clear that the non-linearity can be nullified if $ab = 2 \nu$, leading to a linear heat equation
\begin{equation} \label{heat}
	\partial_t \phi - \nu \; \partial_{xx} \phi = 0.
\end{equation}
In particular, we obtain the Cole-Hopf transformation for $a=1$ and $b=2 \nu$
\begin{equation} \label{ch}
	u = a \partial_x \psi = - 2 \nu \; \frac{\partial_x \phi }{\phi}.
\end{equation}
The initial condition for (\ref{heat}) can be obtained from (\ref{ch}) 
\begin{equation} \label{pic}
	\phi_0(x)=\phi(x,t=0)=\exp \Big( - \frac{1}{2 \nu} \int_0^x u_0(y) \rd y\Big).
\end{equation}
Solutions $\phi$ (and its derivatives) are found by convolution of Green's function and the initial data:
\begin{equation} \label{pgr}
	\phi(x,t) = \int_0^1 G(x,y,t) \; \phi_0(y) \rd y.
\end{equation}
It is evident that (\ref{beqn}) is regular for $t>0$ and categorically rules out blow-up in {\it real} fluids ($\nu >0$) having bounded initial energy.  
\subsection*{Computation of exact solution}
The transformed initial condition (\ref{pic}) can be evaluated without difficulty for $\nu > 0.01$ by means of machine-aided computations as the Fourier series in the definition of the Green's function converges reasonably fast. As the viscosity becomes smaller, it is impractical to compute the analytic expression to acceptable accuracy, because the exponential function in $\phi_0$ would cause underflow or overflow as soon as we integrate away from $x=0$, even over short time intervals. Since velocity $u$ is given by a {\it ratio} of two rapidly-changing exponentials, one remedy is that we scale both $\phi$ and $\phi_x$ by suitably chosen functions so that the numerical difficulty may be avoided. For instance, we can compute
\begin{equation} \label{exsol1}
	\tilde{\phi}_0(x)=\exp(Q) \phi_0(x) 
\end{equation}
and then
\begin{equation} \label{exsol2}
	\tilde{\phi}(x,t) = \int_0^1 G(x{-}y,t) \exp(P) \; \tilde{\phi}_0(y) \rd y,
\end{equation}
where $Q=Q(x,t,\nu)$ and $P=P(x{-}y,t,\nu)$ are the scaling. In practice, it is tedious to find a correct pair, as we must rely on trial and error. Nevertheless, it is clear that the pair are not unique, and depend on the initial data as well as the local flow. The objective of evaluating the exact solutions in this manner is to verify a general-purpose numerical scheme developed in the present paper. In table~\ref{tb:exsol}, we list some results for viscosity down to $10^{-4}$. At least, an order of magnitude improvement has been achieved compared to the best known exact solutions (cf. Cole 1951; Basdevant {\it et al.} 1986; Zhang {\it et al.} 1997). 

\begin{table} 
	\centering
\begin{tabular}{rrr} \hline \hline 
    &  $\;\;\;\;\;\;\nu=10^{-3}$ & $\;\;\;\;\;\;\nu=10^{-4}$   \\ \hline \hline
    $t$ = 0.25 & 13.9337 & 14.5616 \\
    0.5 & 491.9831 &4995.0574   \\ 
		1 & 269.9739 &2710.8183   \\ 
		5 & 17.4706 &176.4549   \\
		10 & 4.5980 &46.8603    \\ 
		100 & 0.0397 & 0.4869    \\ 
		1000 & 0 & 0.0040    \\ \hline \hline
\end{tabular}
\caption{\rm{Evaluation of analytic solution (\ref{ch})-(\ref{pgr}) by the scaling technique. Initial data $u_0(x)=\sin(\pi x)$. Results of $-\partial_x u|_{x=1}$ or $|\partial_x u|_w$ at selected time.}} \label{tb:exsol}
\end{table}
\section{Numerical scheme for small viscosity $\nu$}
We do not need to emphasise the importance of the simplest non-linear differential equation (\ref{beqn}) as a prototype for the physics of fluid dynamics. Yet, there does not seem to exist a reliable computational method at small viscosity $\nu \lesssim 10^{-5}$. Zhang {\it et al.} (1997) give a short review of standard numerical schemes such as finite difference and spectral technique. Unfortunately, none of these numerical methods appears to be capable of handling Burgers' equation in small-viscosity applications; detailed numerical data are scarce in literature. In a nutshell, the difficulty with these methods is their inability to resolve the finest spatio-temporal scales over the transitional regime. Specifically, among a number of published work for high Reynolds-number flows, there are significantly different solution behaviours over $t=0.2$ and $t=0.4$ at $Re=10^5$ for initial data (\ref{bl}). An alternative approach is the mesh-free method of smoothed particle hydrodynamics. By an enhanced formulation, Hashemian \& Shodja (2008) managed to calculate the velocity solutions at specific sets of particle distribution and variable dilation. 
 
Surprisingly, the dynamics defined by (\ref{beqn}) at small finite viscosity {\it is unknown} though efforts made on the theoretical inviscid limit ($\nu=0$) are well-documented. However, it is not the velocity but the vorticity field which enshrines the proliferation process. To explore the shear regime, we develop a numerical scheme which takes advantage of variable mesh distribution and totality of spatio-temporal scales. The first feature allows improved discretisation in shear-intensity regions for high-efficiency while the second is crucial to capture the details of continuum turbulence. The trapezoidal rule is used to approximate kernel convolution in conjunction with an implicit Euler method for time-marching. As the convolution carries over smooth functions, intermediate mesh grids can be obtained by any suitable interpolation procedure, as long as the solutions remain differentiable (analogous to ideas in Adams-Bashforth methods). To be definite, the grid for case $u_0(x)=\sin(\pi x)$ is distributed according to formula
\begin{equation} \label{gride}
	x_i=\frac{1}{2} + \frac{\tanh \big( \sigma(2i{-}n{-}1)/(n{-}1) \big)}{2 \tanh \sigma},
\end{equation}
where the stretching parameter $\sigma$ is used to control the grid spacings near both ends at given grid point $n$. Similar ideas are applied to the shear flow (\ref{shear}), formula,
\begin{equation} \label{gridc}
	x_i=\frac{\tanh \big( 2\sigma (i{-}1)/(n{-}1)\big)}{2\tanh \sigma},
\end{equation}
generates grids in one half of the domain, so as to cluster more points near the centre $x=0.5$. (Either can be enhanced in self-adaptive schemes for calculations of complicated moving shear field.) The approximation error is proportional to $(\Delta t) \: (\Delta x)^2$. It is found that the time step $\Delta t$ has to be one or two orders of magnitude lower than viscosity $\nu$ in order to maintain numerical stability over the phase of non-linear growth. In the initial and decay phases, larger time-steps (one or two orders higher than $\nu$) may be chosen to cut down numbers of iteration. It is a practical matter of trade-off between time $\Delta t$ and  iterations. The majority of our calculations reach a prescribed convergence tolerance ($10^{-14}$) on a Euclidean norm of velocity $u$ in less than $10$ iterations at specified time step. A typical calculation, say from $t=0$ to $t=1$ for $\nu \sim O(10^{-6})$, requires $O(10^7)$ time-steps to complete. The storage requirement is roughly $O(n^2)$ and hence is almost negligible in view of modern computing power. Our numerical scheme demonstrates that accurate solutions of Burgers' equation, down to viscosity one billionth ($\nu=10^{-9}$), can be obtained on a {\ttfamily 64}-bit desktop machine with a quad-core microprocessor. 
\subsection*{Wall shear layer}
Validations of our numerical method are first made with exact solution (\ref{ch})-(\ref{pgr}) computed according to scheme (\ref{exsol1})-(\ref{exsol2}) (see table~\ref{tb:r1k}). As a matter of fact, over 99\% of Burgers' flow evolves significantly in an ultra-thin width at the wall $\sim O(0.005)$ and hence $90\%$ computational effort is allocated to the flow in the vicinity of $x \approx 1$. It is over this dissipative region that the flow possesses scales of various sizes, and nearly consumes all the initial kinetic energy. Even at viscosity $\nu=10^{-3}$, the wall vorticity is increased $160$ times in a time interval of $0.5$. 

\begin{table}
	\centering
\begin{tabular}{ccccc} \hline \hline
    &  $\sigma \;/ \;n$ in (\ref{gride}) & $\Delta t$  & $|\partial_x u|_{\max}$ & $\;\;\;t$ for  $|\partial_x u|_{\max}$  \\ \hline \hline
 Exact &  &   & 495.028422 & 0.503023 \\ \hline
Numerical&4.0 / 121 & $1{\times}10^{-4}$ &  495.35 & 0.50320  \\ 
		  &  4.0 / 121 & $1{\times}10^{-5}$ &  495.93 & 0.50302  \\
      &  4.0 / 161 & $1{\times}10^{-4}$ &  495.24 & 0.50320  \\
      &  4.0 / 161 & $5{\times}10^{-5}$ &  495.11 & 0.50310  \\ 
		  &  4.0 / 161 & $1{\times}10^{-5}$ &  495.40 & 0.50303  \\
		  &  4.0 / 201 & $1{\times}10^{-4}$ &  495.21 & 0.50320  \\ 
		  &  5.0 / 201 & $1{\times}10^{-4}$ &  495.24 & 0.50320  \\ 
		  &  3.0 / 201 & $1{\times}10^{-4}$ &  495.21 & 0.50320  \\  \hline \hline
\end{tabular}
\caption{\rm{Effect of numerical parameters on solution accuracy. Viscosity $\nu=1{\times}10^{-3}$, $u_0(x)=\sin(\pi x)$ with homogeneous boundary condition at $x=0$ and $x=1$. The maximum derivative is found to occur at wall $x{=}1$.  }} \label{tb:r1k}
\end{table}

One crucial aspect in computational fluid dynamics is, how to resolve the broad range of scales in high Reynolds number flow. As the Navier-Stokes equations are globally regular, all of the scales in turbulence must be taken into account on the basis of the continuum (see, for instance, Bradshaw 1971; Tennekes \& Lumley 1972; Davidson 2004). Solutions of Burgers' equation represent the ensemble mean values at space-time location $(x,t)$ for every given finite viscosity. Practically, machine-aided computations are limited by finite precision arithmetic. In real turbulence, there exists a natural cut-off on the flow scales as viscous dissipation cannot be an instantaneous process, and must proceed on finite spatial scales.\footnote{The K41 hypotheses on turbulence (Kolmogorov 1941$a$, 1941$b$) do not apply to our computations as, evidently, the evolved flow field from initial data (\ref{bl}) or (\ref{shear}) cannot be regarded as homogeneous and `isotropic'. Nevertheless, it is interesting to notice that the Kolmogorov dissipation scale for fully-developed turbulence should be $\eta \sim O(\nu^{4/3})$, taking the integral scale $l_0$ as unity. Thus the mean dissipation rate $\varepsilon \sim u^3/l_0 \sim O(1)$ in the present models.} The results given in tables~\ref{tb:scales} and \ref{tb:scales2} show that the spatial resolution in our computations is adequate though somehow excessive. By performing additional numerical experiments, we establish a simple rule of thumb: {\it the minimum cut-off scale has to be at least two orders of magnitude smaller than viscosity}.  
\begin{table}
	\centering
\begin{tabular}{rrrrr} \hline \hline
 Cut-off & $ 10^{-5}\;\;\;$ & $10^{-6}\;\;\;$ & $10^{-7}\;\;\;$ & $\epsilon\;\;\;\;\;\;\;$ \\ \hline
$t=0.3$	 & $37.402\;\;\;$  & $52.130\;\;\;$ &  $54.019\;\;\;$ & $54.220\;\;\;$  \\
$0.32$ & $\;\;\;1.357{\times}10^2$ & $\;\;\;9.602{\times}10^2$ & $\;\;\;1.318{\times}10^3$ & $\;\;\;1.361{\times}10^3$ \\ 
$0.34$ & $8.756{\times}10^3$ & $1.625{\times}10^4$ & $1.700{\times}10^4$ & $1.709{\times}10^4$ \\
$0.4$  & $2.898{\times}10^4$ & $4.022{\times}10^4$ & $4.139{\times}10^4$ & $4.146{\times}10^4$ \\ \hline \hline
\end{tabular}
\caption{\rm{Result of $-\partial_x u|_{w}$ over `transition' regime. Viscosity $\nu=10^{-5}$, $u_0(x)=\sin(\pi x)$. For all the runs, $n = 181$ and time marching $\Delta t =10^{-5}$, and 
$2{\times}10^{-6}$. The symbol $\epsilon \approx 2{\times}10^{-19}$ denotes the cut-off scale used in the present work. Roughly, it equals to a specific floating-point precision implemented by the {\ttfamily GNU FORTRAN 95} compiler. }} \label{tb:scales}
\end{table}

\begin{table}
	\centering
\begin{tabular}{rrrrr} \hline \hline
 Cut-off & $ 10^{-6}\;\;\;$ & $10^{-7}\;\;\;$ & $10^{-8}\;\;\;$ & $\epsilon\;\;\;\;\;\;\;$ \\ \hline
$t=0.2$	 & $4.572\;\;\;$  & $4.580\;\;\;$ &  $4.581\;\;\;$ & $4.581\;\;\;$  \\
$0.3$	 & $52.518\;\;\;$  & $54.634\;\;\;$ &  $54.632\;\;\;$ & $54.652\;\;\;$  \\
$0.32$ & $\;\;\;7.112{\times}10^3$ & $\;\;\;1.517{\times}10^4$ & $\;\;\;1.523{\times}10^4$ & $\;\;\;1.530{\times}10^4$ \\ 
$0.34$ & $1.592{\times}10^5$ & $1.704{\times}10^5$ & $1.714{\times}10^5$ & $1.713{\times}10^5$ \\
$0.4$  & $3.686{\times}10^5$ & $4.099{\times}10^5$ & $4.149{\times}10^5$ & $4.147{\times}10^5$ \\ 
$0.5$  & $3.900{\times}10^5$ & $4.892{\times}10^5$ & $5.007{\times}10^5$ & $5.006{\times}10^5$ \\ 
$0.6$  & $3.241{\times}10^5$ & $4.533{\times}10^5$ & $4.685{\times}10^5$ & $4.691{\times}10^5$ \\ \hline \hline
\end{tabular}
\caption{\rm{See table~\ref{tb:scales}, $\nu=10^{-6}$ and the shortest $\Delta t=2{\times}10^{-7}$. The flow remains `laminar' during the initial phase $0 \leq t < 0.319 $. Scale proliferation starts at $t \approx 0.319$. At $t \approx 0.5$ and onward, the flow in the wall vicinity evolves into the 'fully-developed turbulent' state.  }}  \label{tb:scales2}
\end{table}
\subsection*{Free shear flow}
As we have entered the arena of computations at extremely small viscosity, some efforts have been made to nurse the present low-$\nu$ calculations, particularly on the suitable choice of time-steps $\Delta t$. Should $\Delta t$ be too coarse, there are problems of convergence; too fine, of accumulations of excessive round-off errors. The diverging scenario is easy to diagnose during computations while the latter difficulty is rather tricky to quantify. We are well aware of the fact that it is impossible for machine-aided computations to be immune from round-off. 

The velocity bound (\ref{vbound}) serves as an indication, as the integrations are undertaken from $x=0$ to $x=1$. Reversal of the integrations from $x=1$ to $x=0$ may provide useful hints on the degrees of round-off. Another reason to examine shear flow (\ref{shear}) is its symmetry. As our calculations proceed, we monitor the development of one-direction sums, for instance, the integrated velocity profiles in $0 < x < 0.5 $ and $0.5 < x < 1$ over $10^7$ time steps. Significant departure from velocity symmetry gives us some ideas about the seriousness of accumulating errors. These methods are entirely empirical, and are found to work well for the viscosity range studied in the present paper.

We find that reliable flow solutions for $\nu > 10^{-6}$ can be easily computed in double precision arithmetic. As a practical guide, use of higher precision floating point representation in programming language (for instance, the variable type {\ttfamily real(kind=16)} in modern {\ttfamily FORTRAN}) is probably one of the most effective ways to minimise the round-off errors, assuming the computations may be performed in reasonably manageable time. It is anticipated that accurate solutions of Burgers' equation for any given viscosity value are achievable by the present algorithm on more powerful machines. In practice, a Reynolds number in the order of one trillion $O(10^{12})$ is likely over-specified for the great majority of earth-bound incompressible flows. 
\section{Flow developing stages}
\subsection{Rapid shear build-up}
\begin{figure}[t] \centering
  {\includegraphics[keepaspectratio,height=16.5cm,width=15cm]{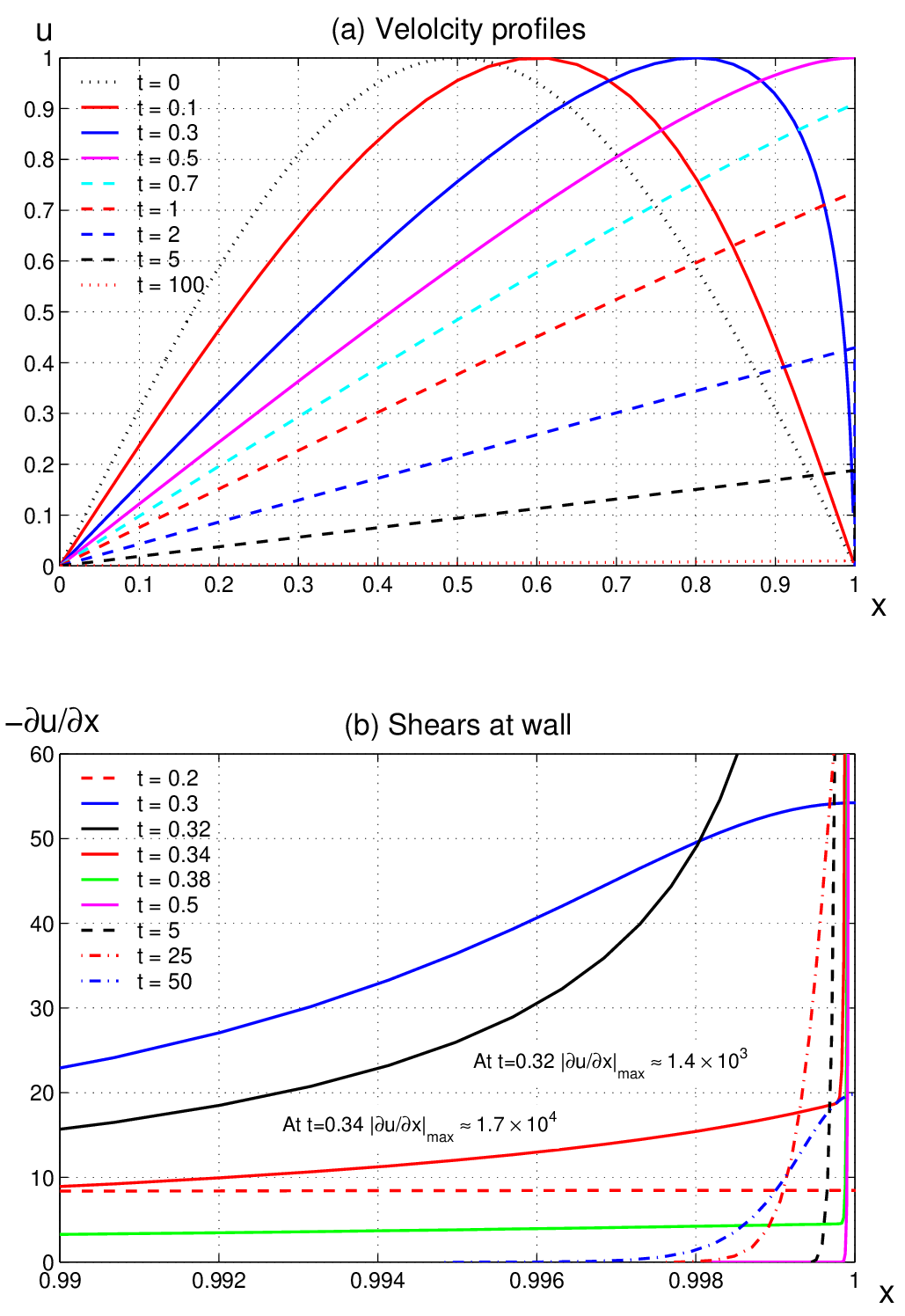}} 
 \caption{Initial velocity $u_0(x) = \sin(\pi x)$, $\nu=10^{-5}$. The calculation was done with a grid distribution of $n=181$ and $\sigma=7$ as defined in (\ref{gride}). To ensure fast convergence, the time-marching was carried out at $\Delta t = 10^{-5}$ to $10^{-6}$. The profile at $t=0.5$ is associated with the maximum shear ($|\partial_x u|_w \approx 5{\times}10^4$). The overall development in velocity profiles (a) looks mundane. But vorticity evolves from initial value $\pi \cos(\pi x)$ and coalesces into a thickness $\sim O(\nu)$ next to the wall, as shown in (b). From $t \approx 0.3$ onward, the growth in $|\partial u/\partial x|$ proceeds in $\Delta t \approx 0.05$ over a mere length $\Delta x \approx 0.001\:$. The viscous wall layer has been well resolved; there are 45 grid points over $0.99 \lesssim x \leq 1$.  } \label{bludu} 
\end{figure}

\begin{figure}[ht] \centering
  {\includegraphics[keepaspectratio,height=16.5cm,width=17cm]{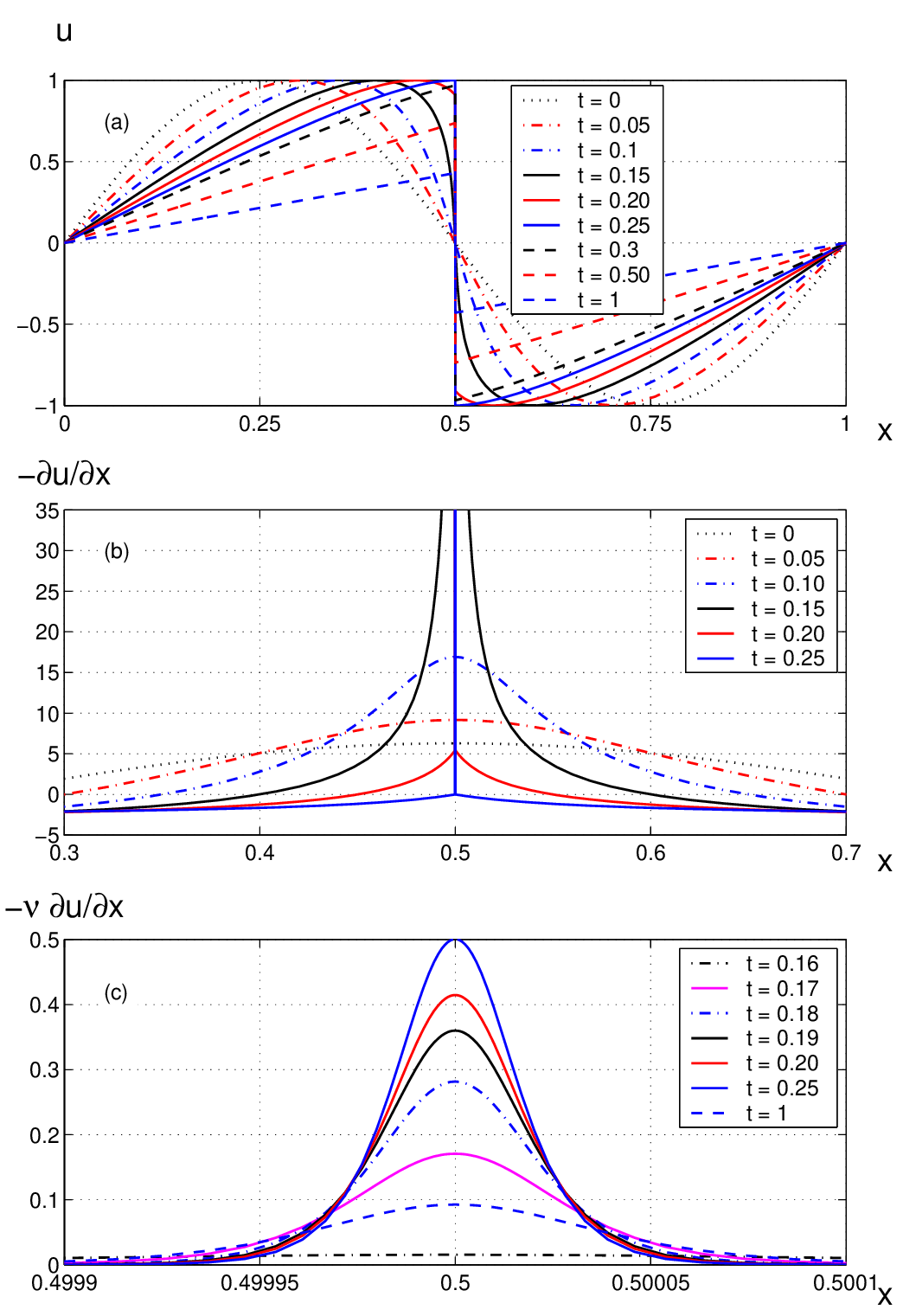}} 
 \caption{Abrupt coalescence of vorticity to form a `shock' in the free-shear layer modelled by initial data $u_0(x)=\sin(2 \pi x)$ at $\nu=10^{-5}$. Velocity profiles in (a) rapidly evolve into a cliff-edge topology in which a large amount of the initial energy clusters. The derivatives show the intensification of the shearing motion at the centre from $t \approx 0.158$ to $0.25$, giving rise to an apparent singularity, see (b). The zoom-in of the coalesced region (c) however reveals the smoothness of the accumulated vorticity. The shock has a finite thickness in which viscous dissipation is intensive. The subsequent decay is somehow extremely slow in the thin layer due mainly to the low viscosity (cf. figure~\ref{bludu}). } \label{slcore} 
\end{figure}

\begin{figure}[ht] \centering
  {\includegraphics[keepaspectratio,height=13cm,width=12cm]{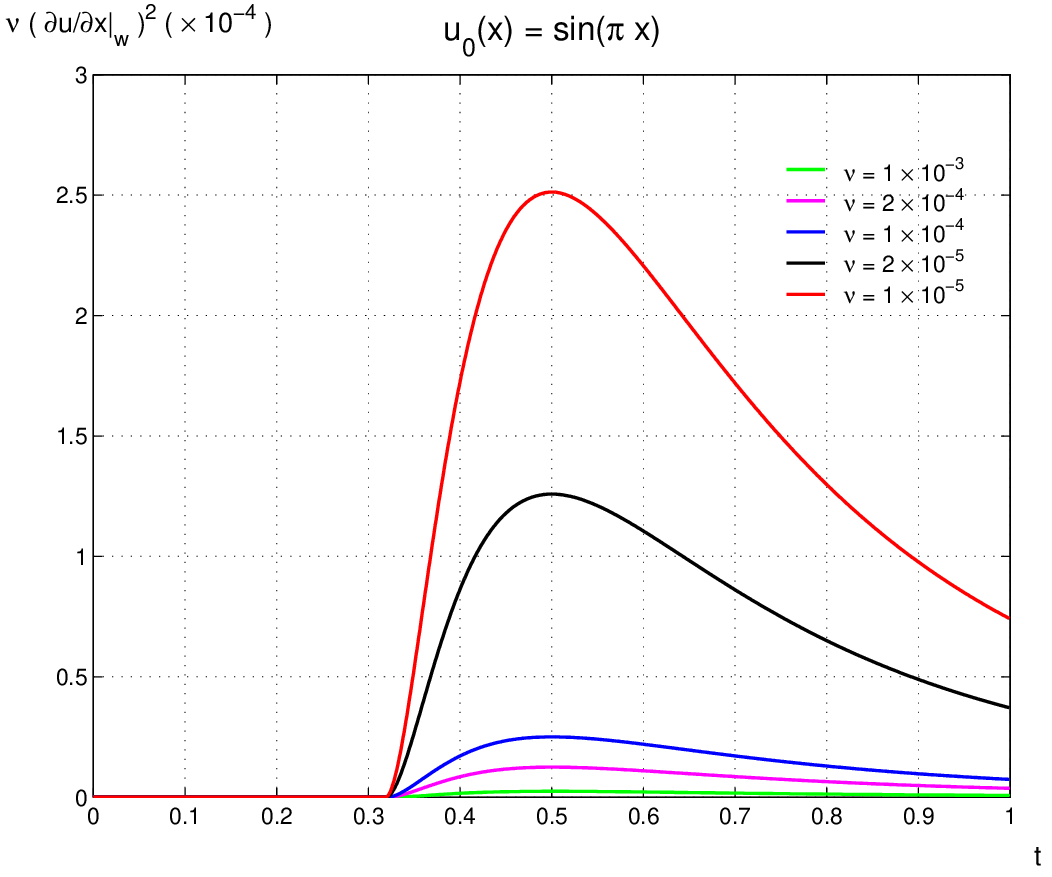}} 
 \caption{The vorticity growth is a consequence of the energy conservation. Evolution of wall enstrophy from the initial vorticity $\partial_x u_0(x) = \pi \cos(\pi x)$ at several viscosity values. The initiation phase of the vorticity build-up depends weakly on viscosity. However, the rapid vorticity increase occurs readily in the flows of low viscosity (cf. typical high Reynolds-number flows).} \label{tevonu} 
\end{figure}

\begin{figure}[ht] \centering
  {\includegraphics[keepaspectratio,height=13cm,width=12cm]{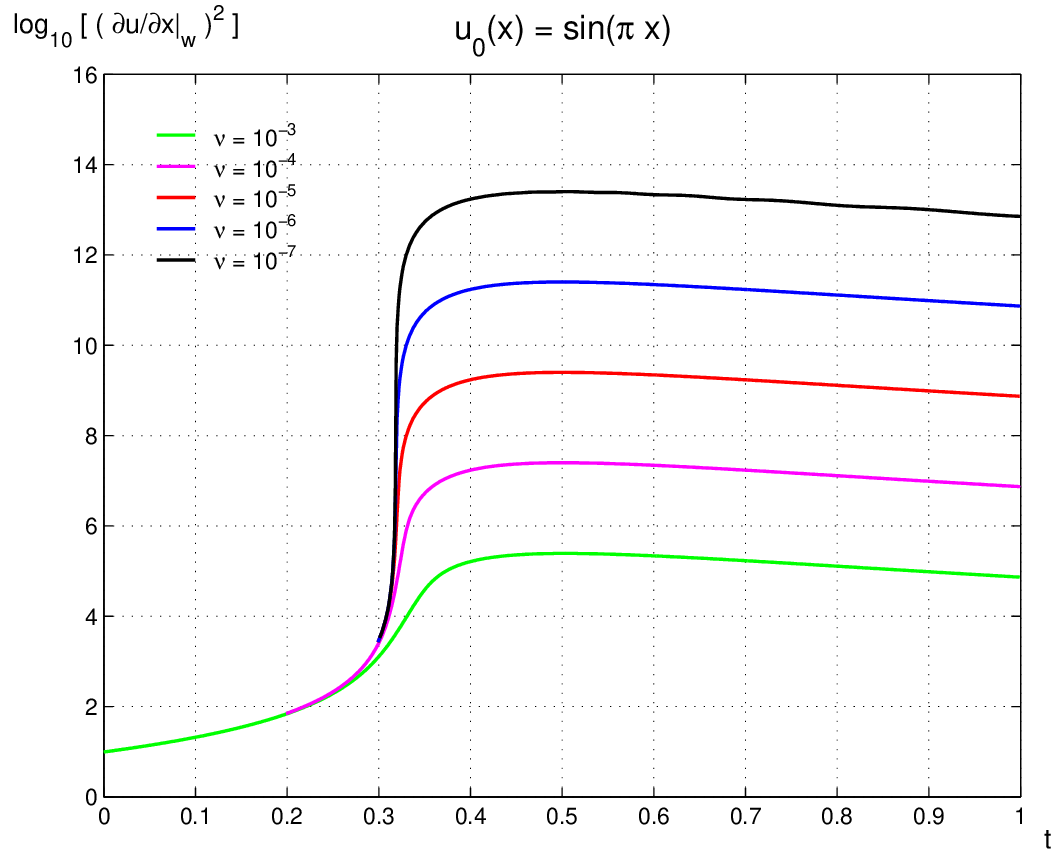}} 
 \caption{Evolution of wall enstrophy as a function of viscosity. At instant $t \approx 0.319$, the wall enstrophy or the vorticity starts to intensify. The build-up is a direct consequence of the scale proliferation by the non-linearity (\ref{ssol}). There are many more vortices of smaller length scales in flows at low viscosity. Prior to the build-up $0< t < 0.319$, the vorticity grows almost linearly: for example, at $t=0.3$, $|\partial_x u|_w=55.05,54.65$ and $54.22$ for viscosity $10^{-7},10^{-6}$ and $10^{-5}$ respectively.  } \label{tevolonu} 
\end{figure}

\begin{figure}[ht] \centering
  {\includegraphics[keepaspectratio,height=17.5cm,width=16cm]{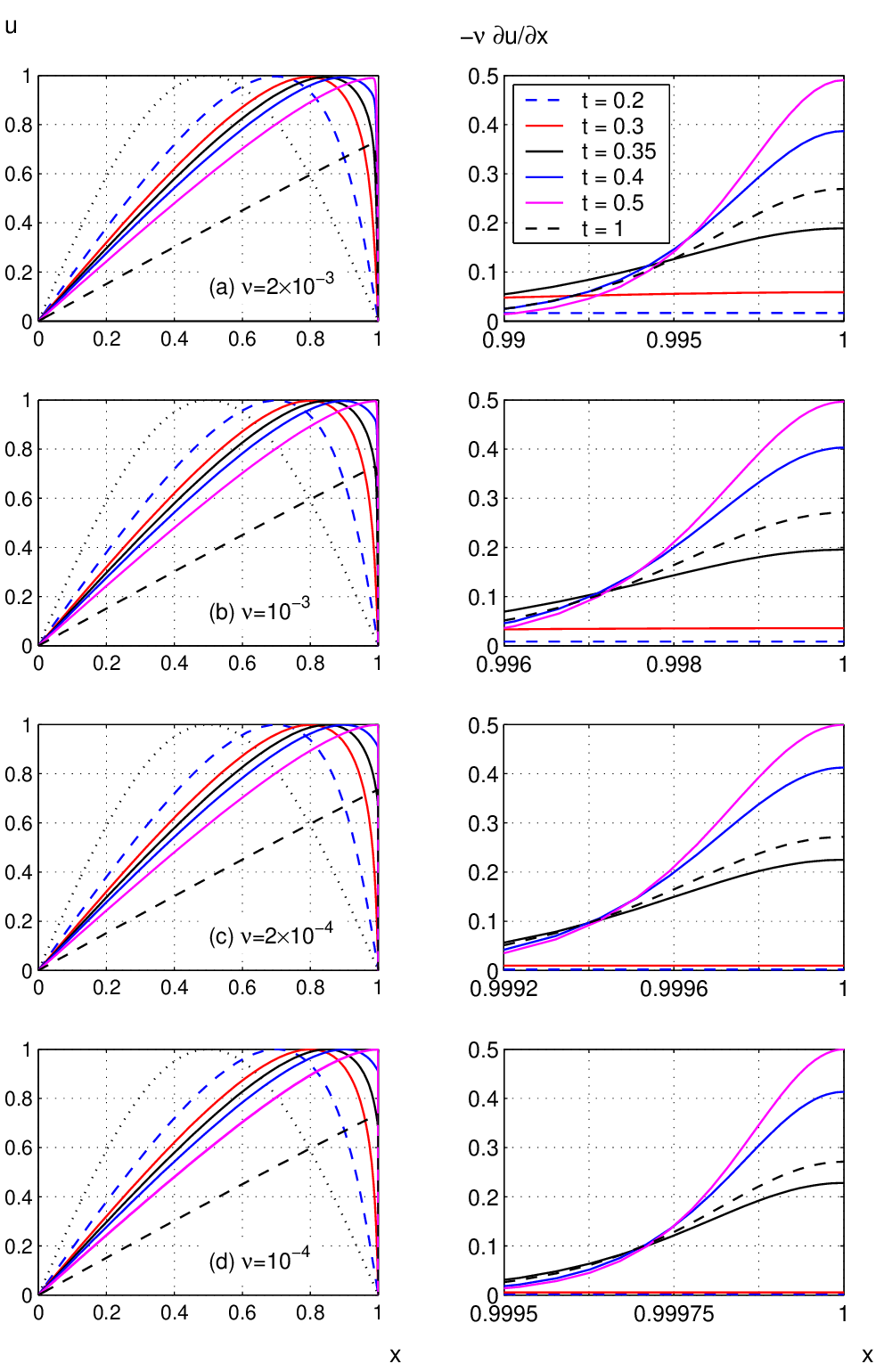}} 
 \caption{At large viscosity, $\nu > 1{\times}10^{-3}$ for $u_0(x) = \sin(\pi x)$, the increase in shears from t = 0.3 to t~=~0.35 is relatively weak. In (a) and (b), the profiles between $0.3 \leq t \leq 0.5$ are not fully coalesced into the narrow wall region. } \label{re4s} 
\end{figure}

\begin{figure}[ht] \centering
  {\includegraphics[keepaspectratio,height=17.5cm,width=16cm]{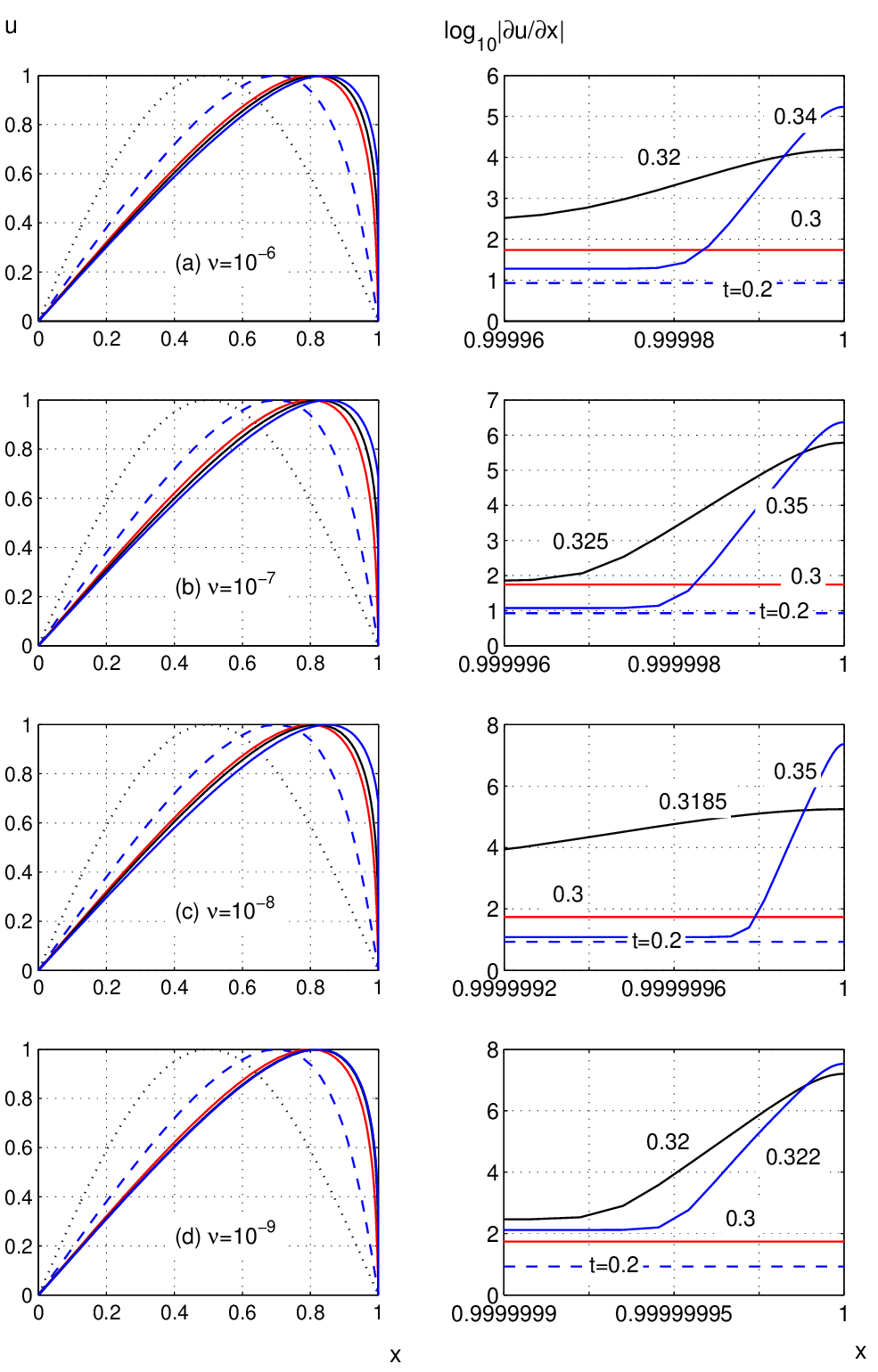}} 
 \caption{Effect of viscosity up to $\nu=10^{-9}$. From the top to the bottom row, grid meshes $(n,\sigma)$ are $(201,8)$, $(211,9)$, $(231,10)$ and $(251,11)$ respectively. The general trend of the shear build-up follows that of higher viscosity, but the upsurges occur more rapidly. The wall vorticity is multiplied by several orders of magnitude in an interval $\Delta t \sim O(0.005)$. However, the modification in the velocity profiles seems insignificant. In terms of practical time and length scales, these flows are easily perceived as a `broken-down' state. } \label{re4u} 
\end{figure}

\begin{figure}[ht] \centering
  {\includegraphics[keepaspectratio,height=17cm,width=17cm]{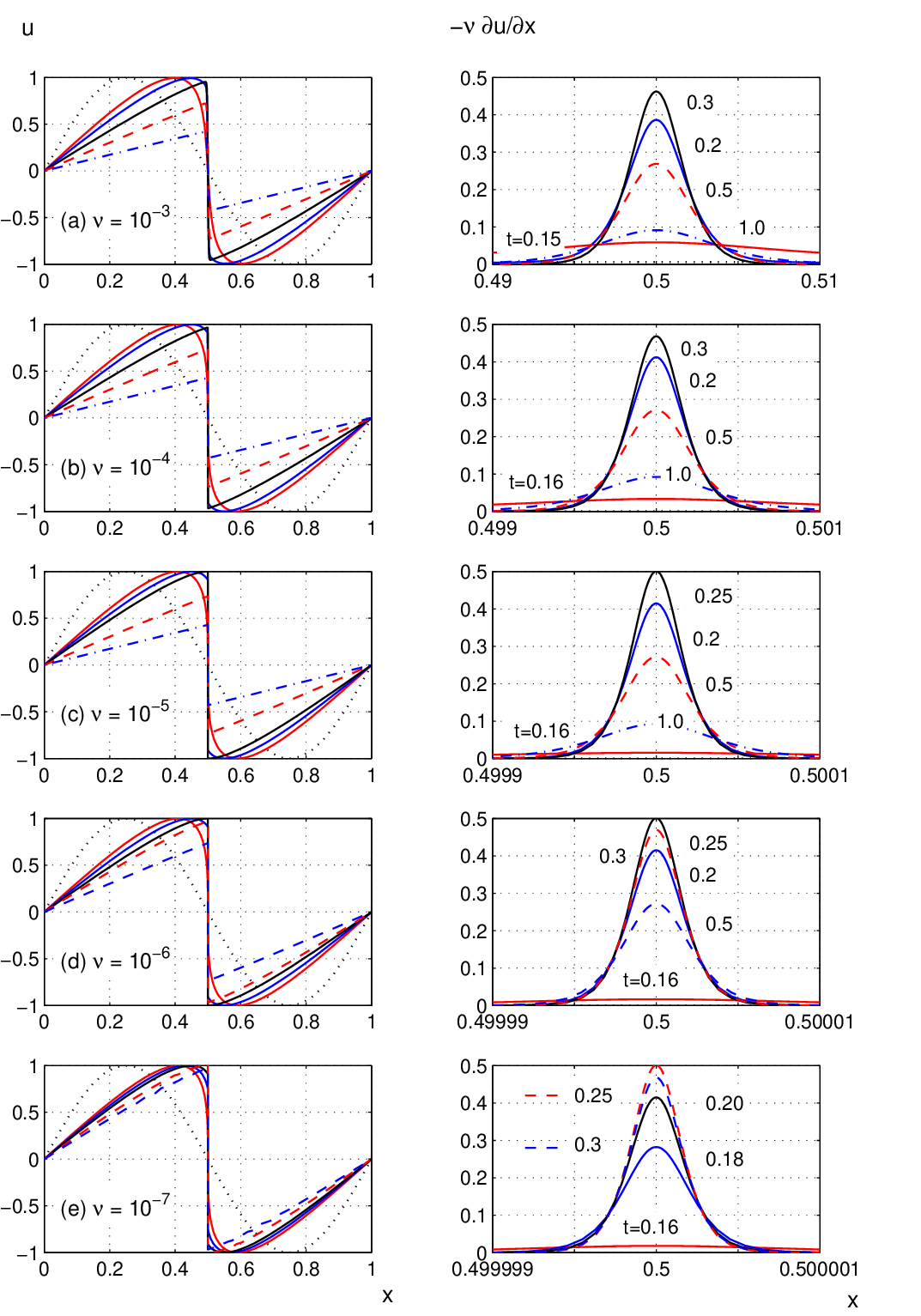}} 
 \caption{Velocity and vorticity results at various viscosity. Initial data $u_0(x)=\sin(2 \pi x)$. The temporal resolutions are in the order of $\nu$. In each zoom-in view of the right column, there is a sharp increase in the local $\partial u/\partial x$ from time $0.158$ to $0.25$ (the transition period). The velocity profiles of the left column display similar evolution characters largely independent of viscosity (for the identical initial data).} \label{effnu} 
\end{figure}

\begin{figure}[ht] \centering
  {\includegraphics[keepaspectratio,height=18cm,width=15cm]{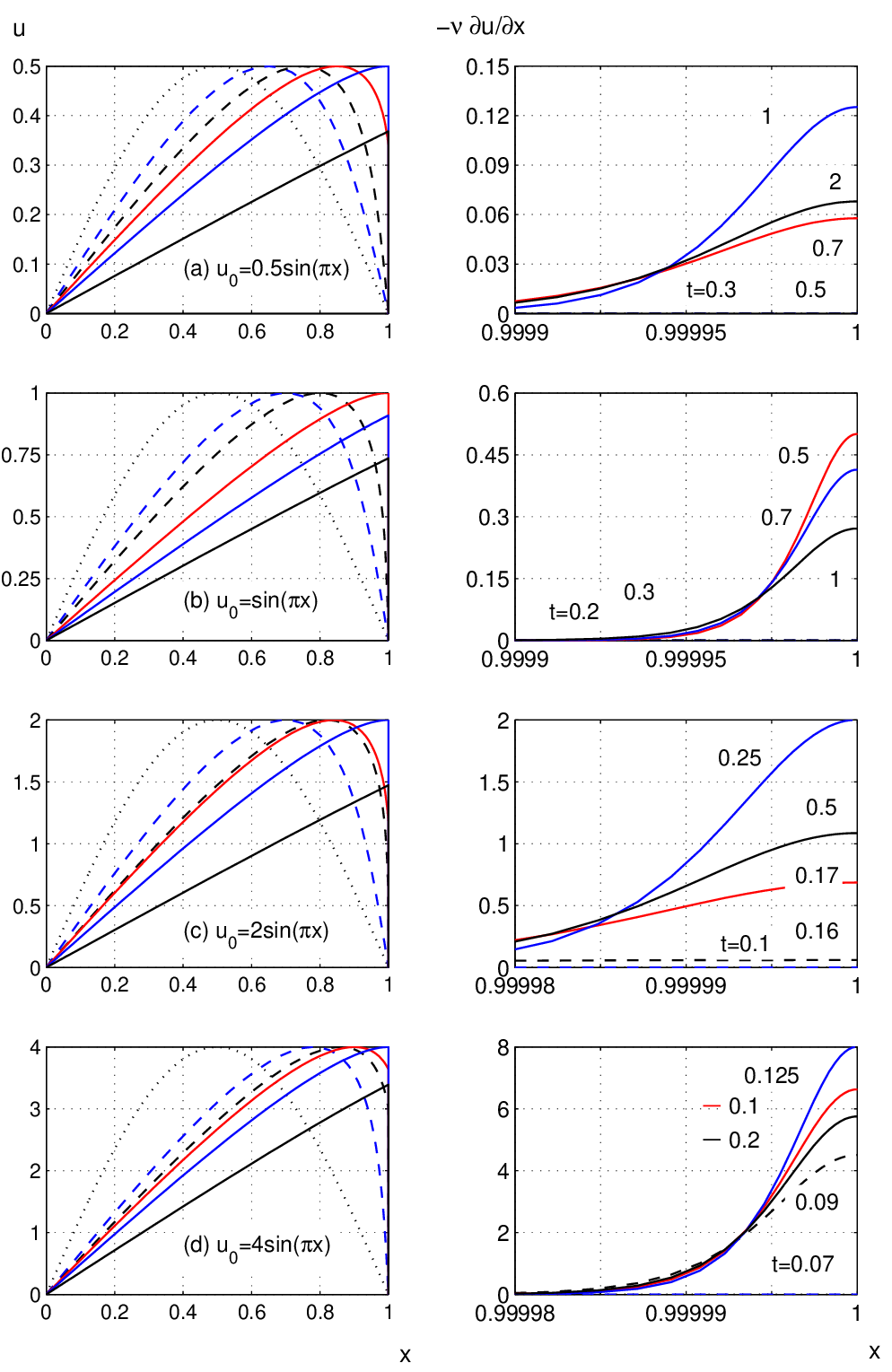}} 
 \caption{Effect of initial strength, $\nu=10^{-5}$. The intensification begins sooner over narrower space as the initial strength becomes stronger. As labelled, the small-time cases in the right column appear as horizontal lines, because their intensities are too weak to be significant on the plotting scales. } \label{rkamp} 
\end{figure}

\begin{figure}[ht] \centering
  {\includegraphics[keepaspectratio,height=18cm,width=17cm]{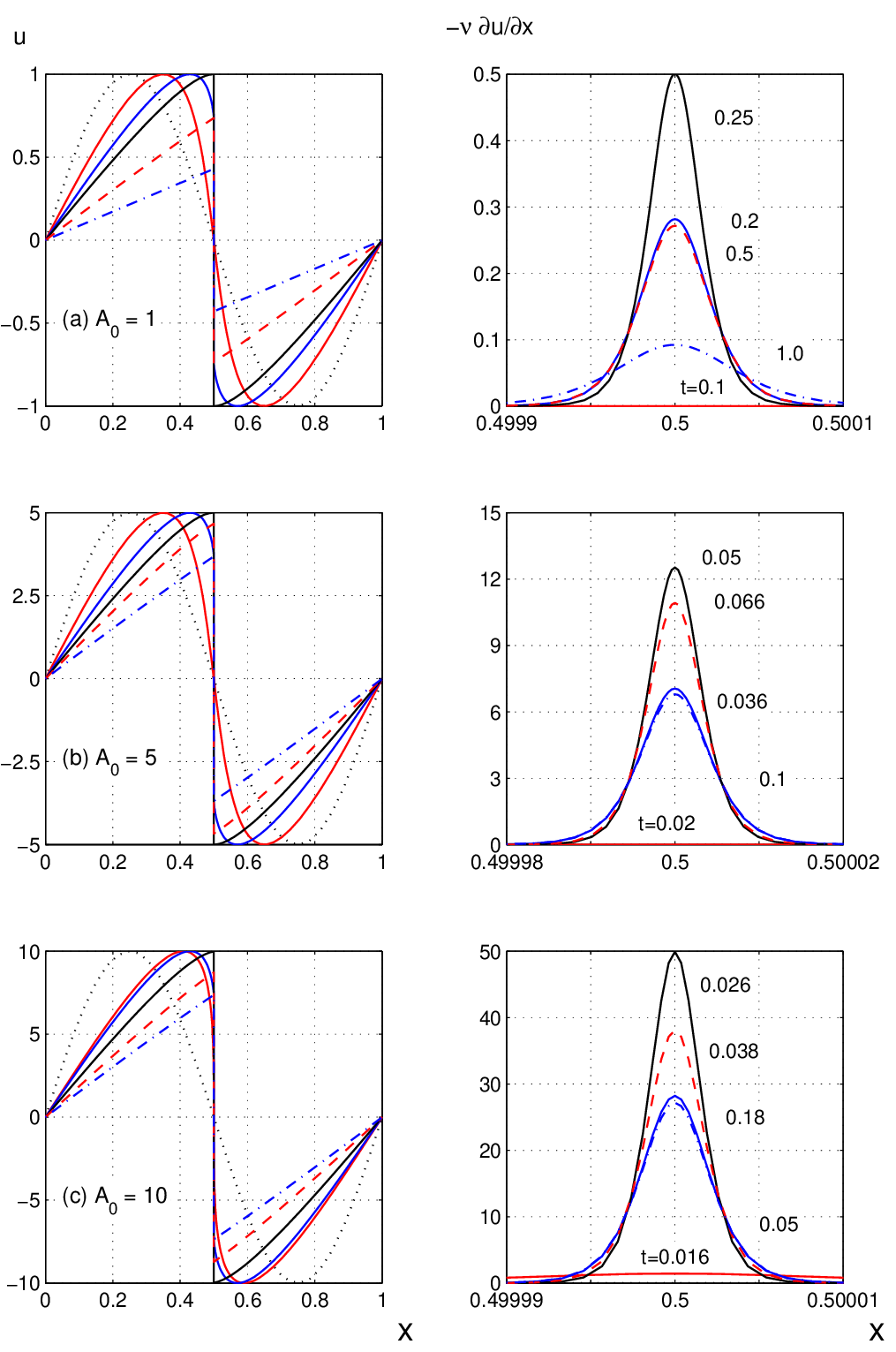}} 
 \caption{Effect of data size $u_0(x)= A_0 \sin(2 \pi x), \;\nu=10^{-5}$. The shear scaled by viscosity on the right equals the `shear stress' per unit energy, similar to 
	(\ref{wallshear}). It may be normalised by $u_0^2/2$ so that, in all the three cases, the maximum shear stress is in the order of unity.} \label{effamp} 
\end{figure}

In table~\ref{tb:r10k}, we compare our numerical solutions with exact values for the case of $\nu=10^{-4}$. Also included are the results of Hashemian \& Shodja (2008). 
\begin{table}
	\centering
\begin{tabular}{r|ccc|rr} \hline \hline
 $x$   &  $\;u$ (Exact) & $u$ (Num.) & $u$ (H-S 2008) & $ \;\partial_x u$ (Exact)& $ \partial_x u$ (Num.) \\ \hline \hline
     0.05 &  0.03792 & 0.03792 &   0.0379 &  0.75837 &    0.75838  \\
     0.11 &  0.08341 & 0.08341 &   0.0834 &  0.75786 &    0.75787  \\
     0.16 &  0.12129 & 0.12129 &   0.1213 &  0.75714 &    0.75715  \\
     0.22 &  0.16668 & 0.16668 &   0.1667 &  0.75591 &    0.75592  \\
     0.27 &  0.20444 & 0.20445 &   0.2044 &  0.75457 &    0.75457  \\
     0.33 &  0.24966 & 0.24966 &   0.2497 &  0.75255 &    0.75256  \\
     0.38 &  0.28724 & 0.28724 &   0.2872 &  0.75052 &    0.75053  \\
     0.44 &  0.33219 & 0.33219 &   0.3322 &  0.74762 &    0.74763  \\
     0.50 &  0.37694 & 0.37695 &   0.3769 &  0.74417 &    0.74418  \\
     0.55 &  0.41407 & 0.41407 &   0.4141 &  0.74084 &    0.74085  \\
     0.61 &  0.45839 & 0.45839 &   0.4584 &  0.73622 &    0.73622  \\
     0.66 &  0.49509 & 0.49509 &   0.4951 &  0.73179 &    0.73179  \\
     0.72 &  0.53882 & 0.53882 &   0.5388 &  0.72568 &    0.72568  \\
     0.77 &  0.57496 & 0.57496 &   0.5749 &  0.71982 &    0.71982  \\
     0.83 &  0.61791 & 0.61791 &   0.6179 &  0.71170 &    0.71170  \\
     0.88 &  0.65330 & 0.65331 &   0.6533 &  0.70388 &    0.70387  \\
     0.94 &  0.69522 & 0.69522 &   0.6952 &  0.69291 &    0.69291  \\
     0.96 &  0.70903 & 0.70904 &   0.7090 &  0.68882 &    0.68881  \\
     0.98 &  0.72277 & 0.72277 &   0.7228 &  0.68446 &    0.68446  \\
     0.99 &  0.72960 & 0.72960 &   0.7296 &  0.68219 &    0.68218  \\
     0.995 & 0.73301 &  0.73301 &   0.7330 &  0.68102 &    0.68101  \\
     0.999 &  0.73480 & 0.73482 &   0.7348 &  -6.18320 &   -6.18259*  \\
     0.9995 & 0.69991 & 0.69998 &   0.6999 &   -259.09 & -259.05  \\ 
     0.9999 &  0.25946 & 0.25955 &   0.2599 &  -2374.05 & -2374.45  \\ \hline
		 1.0  &  0  & $O(10^{-20})$   &  -  & -2710.82 & -2712.01 \\ \hline \hline
\end{tabular}
\caption{\rm{Solutions of Burgers' equation at $t{=}1$ for $A_0{=}1$ in initial profile (\ref{bl}), $\nu{=}10^{-4}$. The exact values are obtained by the proposed procedure (\ref{exsol1})-(\ref{exsol2}). The present numerical calculations (Num.) are carried out with grid points $n=161,181,201,221,241,261$, and $\sigma=5$ at time steps $\Delta t = 1 {\times}10^{-5}$ and $5{\times}10^{-6}$. Over the region $0.99 \lesssim x \leq 1$, we have collocated $45,50,55,60,65,70$ grid points respectively. The results of our computations are found to be independent of the grids. The noticeable discrepancy is only in the last three $x$ locations; these data agree within a relative band of $\pm 0.02\%$. Our data are the values at $n=201$. (*This value is the only exception; it is extrapolated from the runs.) In particular, the wall shear $-\nu \partial_x u|_w$ is found to be $0.2712\pm0.0003$.} } \label{tb:r10k}
\end{table}
In figures~\ref{bludu} and \ref{slcore}, we display our computational results for $\nu=10^{-5}$. The effect of viscosity is examined in figures~\ref{tevonu} and \ref{tevolonu}. 
Detailed local vorticity intensifications are given in figures~\ref{re4s}-\ref{re4u}. The snapshots on the right columns show how the wall shears grow - a property well-established in experiment on turbulent shear flows (see, for instance, \S Turbulence intensity and shearing stress in Klebanoff 1955).
The rapid accumulation of vorticity is not restricted to the wall region. Figure~\ref{effnu} demonstrates the formation of shock waves, which are intensified viscous shears having finite thickness. The size of the initial data determines the initial moment of the abrupt vorticity build-up (figures~\ref{rkamp}-\ref{effamp}).
\subsection{Overshoot in local enstrophy}
\begin{figure}[ht] \centering
  {\includegraphics[keepaspectratio,height=9.5cm,width=15cm]{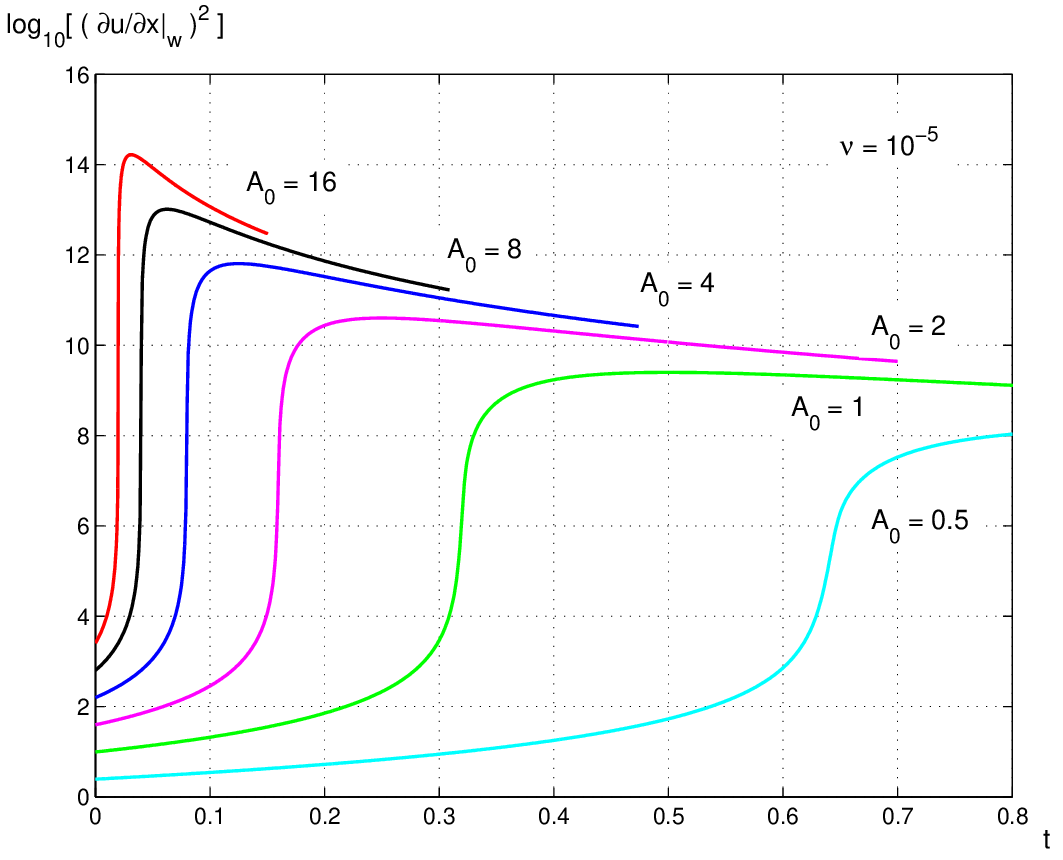}} 
 \caption{Evolution of wall enstrophy as a function of data size, $u_0(x) = A_0 \sin(\pi x)$. For sufficiently large data, there is an {\it overshoot} just after the intensive non-linear growth, corresponding to a temporary increase in the local turbulent skin friction coefficient. The overshoot phenomenon depends on the size of initial data. As the initial data increase, the initial laminar runs diminish, or may even disappear completely. (In higher space dimensions, the vertical jumps are expected to reduce, owning to the extra degrees of freedom.) These `step-ramps' are the trademark of the laminar-turbulent transition. In the present figure, time $t$ may be interpreted as a measure of the Reynolds number as $R_x=x u/\nu \sim t u^2/\nu \propto t$ in time-mean flow. From wind-tunnel experiments, it is known that the local skin friction of turbulent boundary layer settles down on a characteristic function of the Reynolds number (cf. Fig. II.25 of Lighthill 1963).} \label{tevoamp} 
\end{figure}

\begin{figure}[ht] \centering
  {\includegraphics[keepaspectratio,height=17cm,width=15cm]{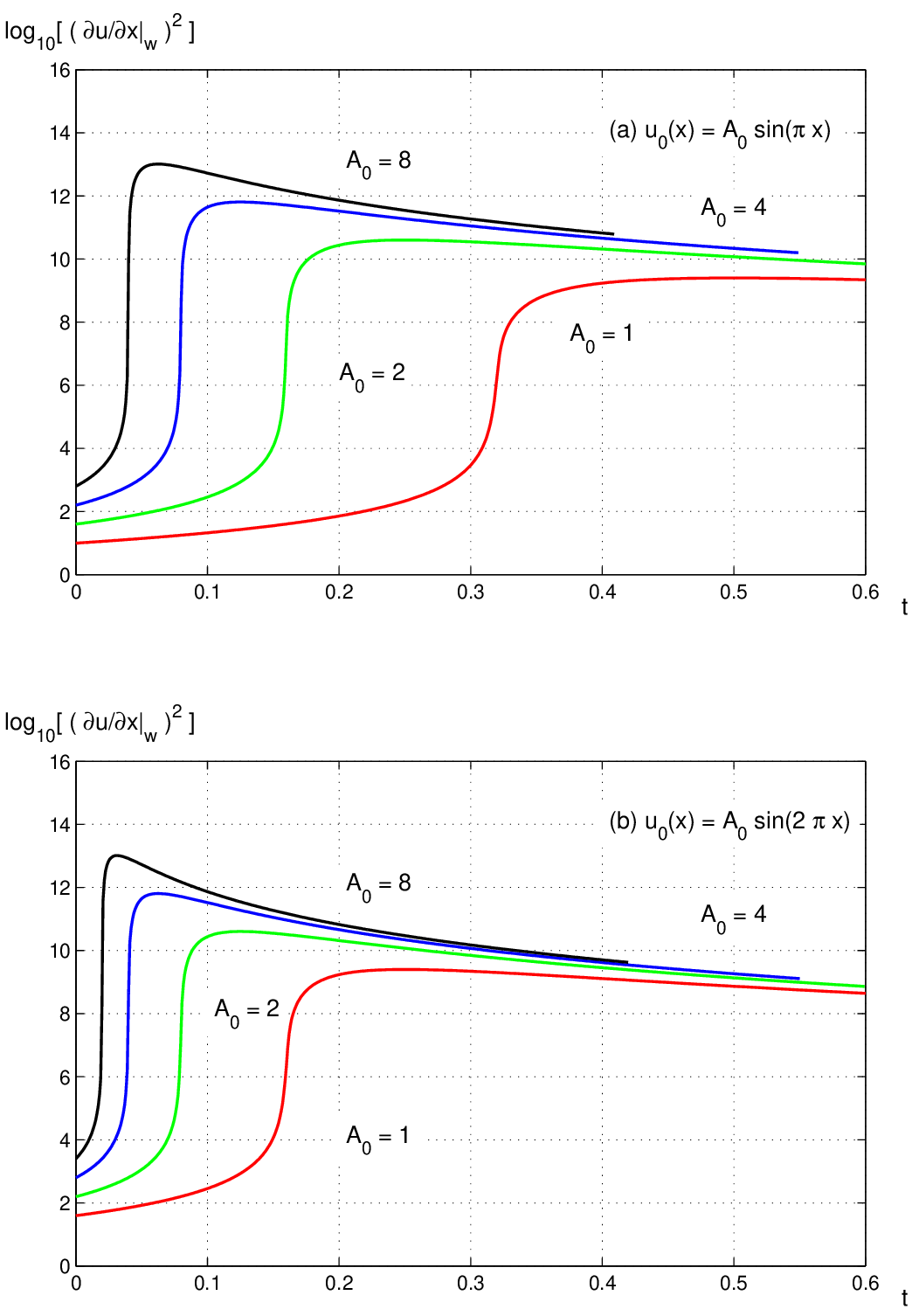}} 
 \caption{Overshoot in local enstrophy as a genuine feature during flow development. Viscosity $\nu=10^{-5}$. There are well-defined overshoots in either case, particularly in flows of large initial amplitudes. Note that either group of the vorticity soon converges to an identical decay, regardless of the initial size. } \label{tevocmp} 
\end{figure}

\begin{figure}[ht] \centering
  {\includegraphics[keepaspectratio,height=17cm,width=15cm]{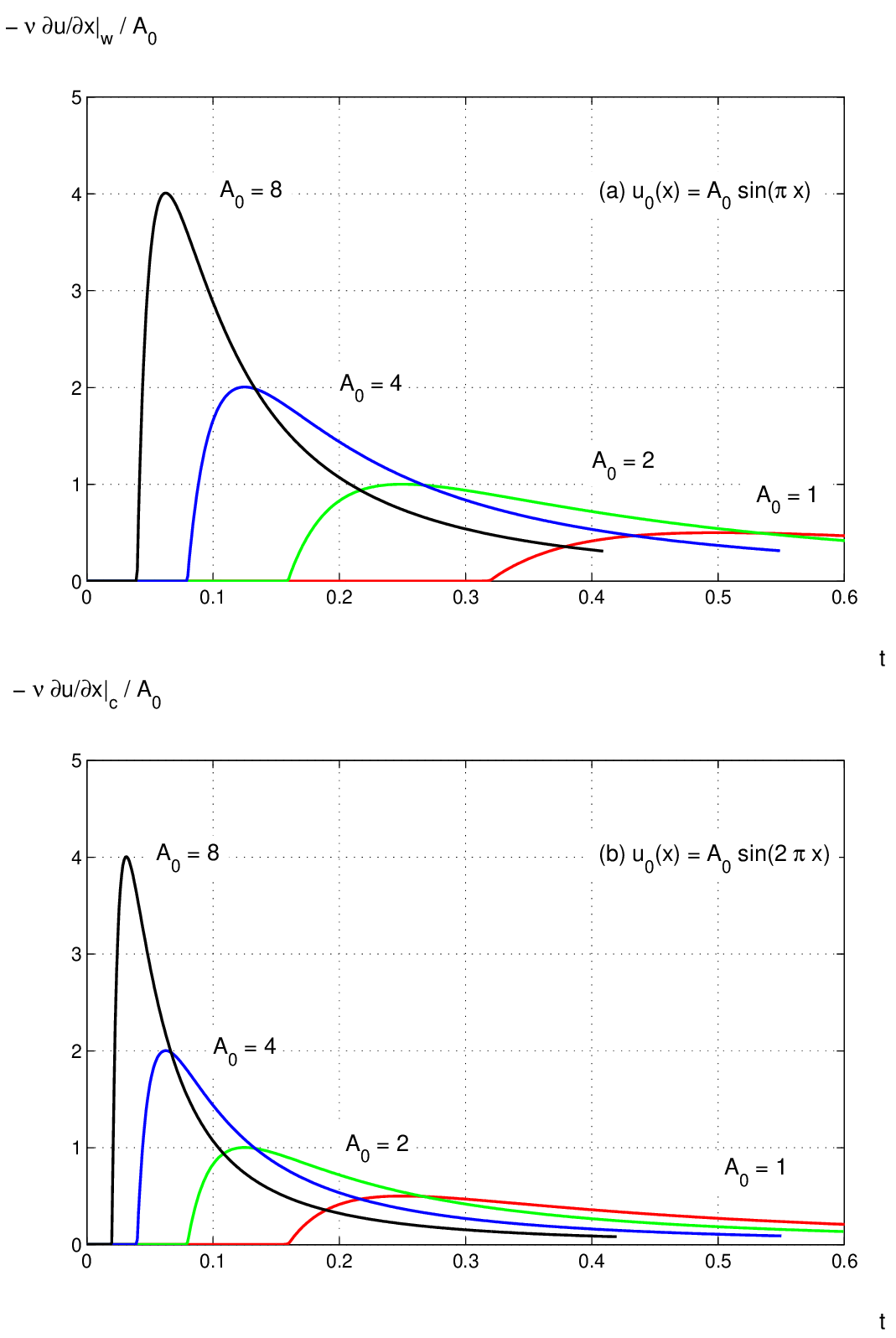}} 
 \caption{Maximum shear as a function of time, $\nu=10^{-5}$. Over the initial phase (laminar state), the magnitudes of vorticity are increasing with time, but they are too small to show up in the vertical scale. In each case of (a), the intensified shear over the abrupt jump resembles the increased skin-friction in a typical turbulent flow through the process of the laminar-turbulent transition. Each accumulated shear in (b) has a steeper rise, or drop. (In higher space dimensions, this property likely contributes to the diverse topologies of vortices commonly existed in high-Reynolds-number free-shear flows.) } \label{tevocf} 
\end{figure}

We have mentioned that the vorticity intensification is a direct consequence of the non-linearity in the equation. Our solutions show that the development enstrophy in Burgers' equation actually reaches a maximum, depending on the strength of the initial data. The phenomenon of `overshoot' in local skin friction was firstly established in boundary layer experiments, see, for example, Coles (1954). In figures~\ref{tevoamp} and \ref{tevocmp}, we present selected computations which verify the existence of the overshoot in our continuum fluid model. The results are given in terms of the local enstrophy, as it is also a quantity applicable to shear layers as well. For the sake of reference, we plot the local `skin friction' coefficient in figure~\ref{tevocf}.

In terms of the intermittency distribution of turbulent spots collected from experiments, Narasimha (1984) gave an empirical explanation of the skin-friction overshoot. In the framework of boundary layer approximations, the overshoot was further explained, and was attributed to an origin shift of the developing turbulent boundary layer over the transition zone (p. 385 of Narasimha 2011). For some practical purposes, the fitted theory may be adequate. The overshoot phenomenon, which was often discounted (p. 639 of Schlichting 1979), is in fact an evolution phase in any fluid motion undergoing the laminar-turbulent transition, particularly the natural one. If we are dealing with the full equations of motion, the concept of a boundary layer becomes somehow blurred. The present computational results reveal that {\it the overshoot in the local enstrophy is a characteristic of the non-linear vorticity build-up over the transition period}.
\subsection{Long time decay}
\begin{figure}[t] \centering
  {\includegraphics[keepaspectratio,height=9.5cm,width=15cm]{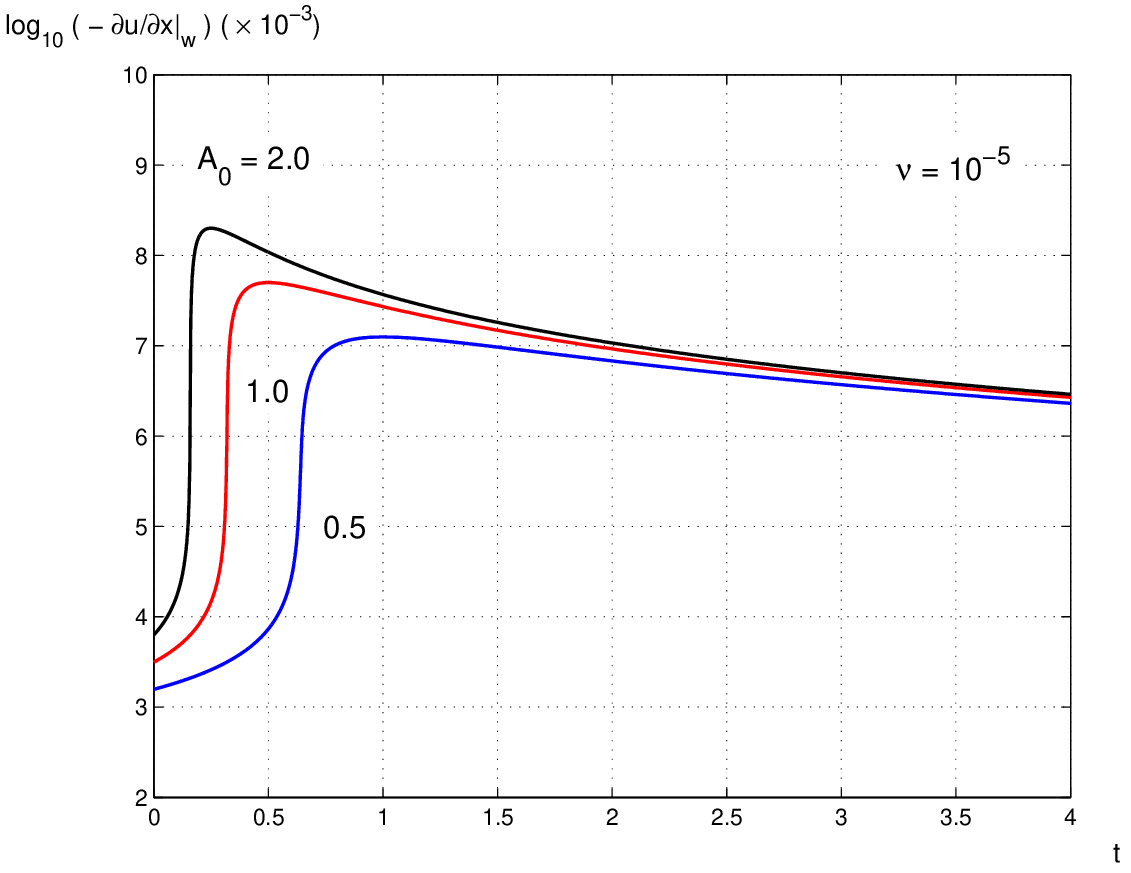}} 
 \caption{Time decay of wall shear, $u_0(x) = A_0 \sin(\pi x)$. The time scale $t \sim O(1)$ covers the initial period in typical applications. The rate of the wall enstrophy decay flattens out after $t > 4$, and indeed the turbulent flows are largely history-independent as advocated in theory. Because of the transition, the short-time impact due to the initial data must be consequential in flow controls (see also figure~\ref{tevocmp}). } \label{tevot} 
\end{figure}

\begin{figure}[ht] \centering
  {\includegraphics[keepaspectratio,height=17cm,width=15cm]{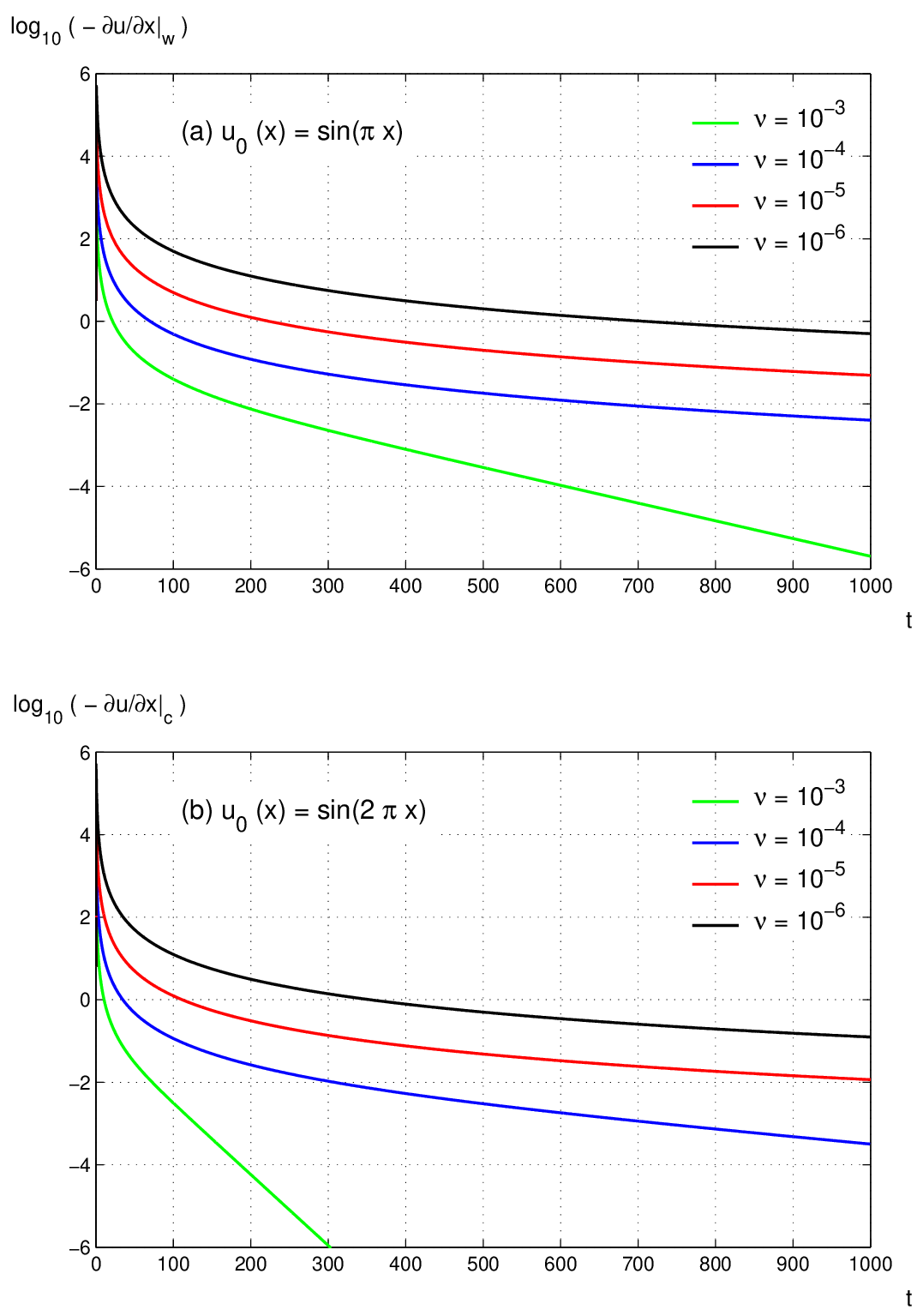}} 
 \caption{Decay of Burgers turbulence at various values of viscosity. The long-term decay is largely identical for the two cases where the initial energies are in the same order of magnitude. The trend in the green lines may be related to the decay of the laminar flow.} \label{tevoblsh} 
\end{figure}

The energy conservation suggests that strong dissipation exists over the thin viscous layer over a long period of time, and thus the velocity undergoes sustained decay. Table~\ref{tb:decay} shows the decay of the wall vorticity.
\begin{table}
	\centering
\begin{tabular}{rc} \hline \hline
 $t\;\;\;$ & $ -\partial_x u|_{w}$ \\ \hline
$223$ & $1$ \\
$703 $ & $10^{-1}$ \\
$2190$	 & $10^{-2}$    \\
$6597$ & $10^{-3}$  \\ 
$17925$ & $10^{-4}$  \\
$37745$  & $10^{-5}$  \\ 
$60605$  & $10^{-6}$  \\ \hline \hline
\end{tabular}
\caption{\rm{Decay of wall shear as time $t \rightarrow \infty$. Data $u_0(x){=}\sin(\pi x)$ and $\nu{=}10^{-5}$. The large time stated on the left below the third row is for indicative purposes. As a rough guide, the maximum velocity drops below $10^{-5}$ after $t > 10^5$. }} \label{tb:decay}
\end{table}
Figure~\ref{tevot} shows that the turbulence quickly forgets its initial energy content at small viscosity. The characteristic of the long-time decay is presented in figure~\ref{tevoblsh}. For $\nu < 10^{-5}\;(A_0=1)$, the local enstrophy attenuates according to  
\begin{equation*}
	\partial u/\partial x \big|_w \; \sim  \; t^{-2.05}\;\;\;\;\; \mbox{as} \;\;\;\;\; t \rightarrow \infty
\end{equation*}
in the boundary layer starting from (\ref{bl}), and 
\begin{equation*}
	\partial u/\partial x \big|_c \; \sim \; t^{-2}\;\;\;\;\; \mbox{as} \;\;\;\;\; t \rightarrow \infty
\end{equation*}
at the location of the maximum viscous dissipation in the turbulent free-shear originated from data (\ref{shear}). The decay graphs in figure~\ref{tevot} assert that the initial data (known as large eddies) have no influence on the attenuation rate. Hence these decay laws are valid for arbitrary finite $A_0$. 

Since the vorticity near the wall, or around the centre, accounts for over $95\%$ of the total shear strength, the local decays suggest that the law of energy decay is
\begin{equation*}
	\int_0^1 u^2(x,t) \rd x \; \sim \; t^{-1} \;\;\;\;\; \mbox{as} \;\;\;\;\; t \rightarrow \infty.
\end{equation*}
\section{Experiments in flow transition}
Our numerical results may be compared to experiment, particularly to the case of Blasius boundary layer on a flat plate. To relate the present calculations to the experiments, we may take the length scale $x=1$ {\ttfamily m}  and the initial speed $|u_0|=1$ {\ttfamily m/s}. The phenomenon of the shear intensification takes place in a time $t \sim O(10^{-3})$ {\ttfamily s} in fluids with viscosity $\nu = 10^{-5}$ {\ttfamily m$^{2}$/s}. 

\begin{figure}[ht] \centering
  {\includegraphics[keepaspectratio,height=10.5cm,width=15cm]{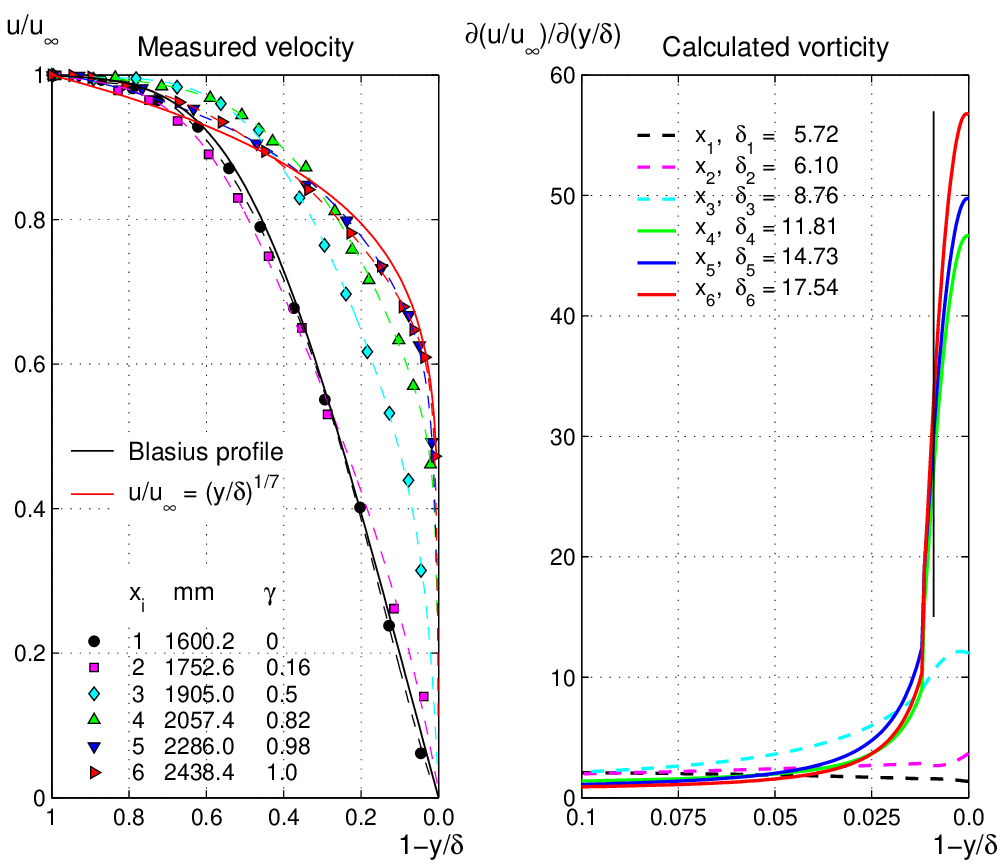}} 
 \caption{Measured velocity at $6$ streamwise-stations over the laminar-turbulent transition in a flat-plate boundary layer as presented in figure~3 of Schubauer \& Klebanoff (1955). Symbols are the experimental data, and lines are best fitted rational functions. The small waviness in the calculated vorticity is largely due to the `read-out' errors from the sparse data points. Position $x$ is the distance from the leading edge and the values given in the legend are estimated boundary layer thickness $\delta$ (in {\ttfamily mm}). The uniform free-stream speed was $u_{\infty}=24.38$ {\ttfamily m/s} and the free-stream turbulence level $0.03\%$. The transition zone covered a length $\sim 700$ {\ttfamily mm}. The tests were carried out at normal room conditions (kinematic viscosity $\nu \approx 1.5{\times}10^{-5}$ {\ttfamily m$^2$/s}). The intermittency factor $\gamma$ was the measured fraction of the total time that the local flow was turbulent. It took merely $30$ {\ttfamily ms} for the (non-dimensional) vorticity to grow $15-20$ times from the Blasius state $x_2$ to station $x_6$, taking into account the uncertainty in the wall profile measurements across the transition region; the abrupt shear increase was squeezed into a thin layer of $5$ {\ttfamily mm} above the wall ($\delta_4$ to $\delta_6$). The unresolved area is indicated by a vertical black line in the vorticity plot, corresponding to the first measurement point from the wall. The peak velocity fluctuations were measured in a width of $1.5$ {\ttfamily mm} next to the wall (cf. figure~4 of the test data). Based on the theory presented in fig. II. 25 of Lighthill (1963), the jump in local skin friction $C_f$ at transitional Reynolds number $3.3{\times}10^6$ is $\Delta C_f \approx 0.003$ (including the local overshoot) which translates into a vorticity increase of $40$ at the end of transition $x_6$.  } \label{skexpt} 
\end{figure}

\begin{figure}[ht] \centering
  {\includegraphics[keepaspectratio,height=10cm,width=15cm]{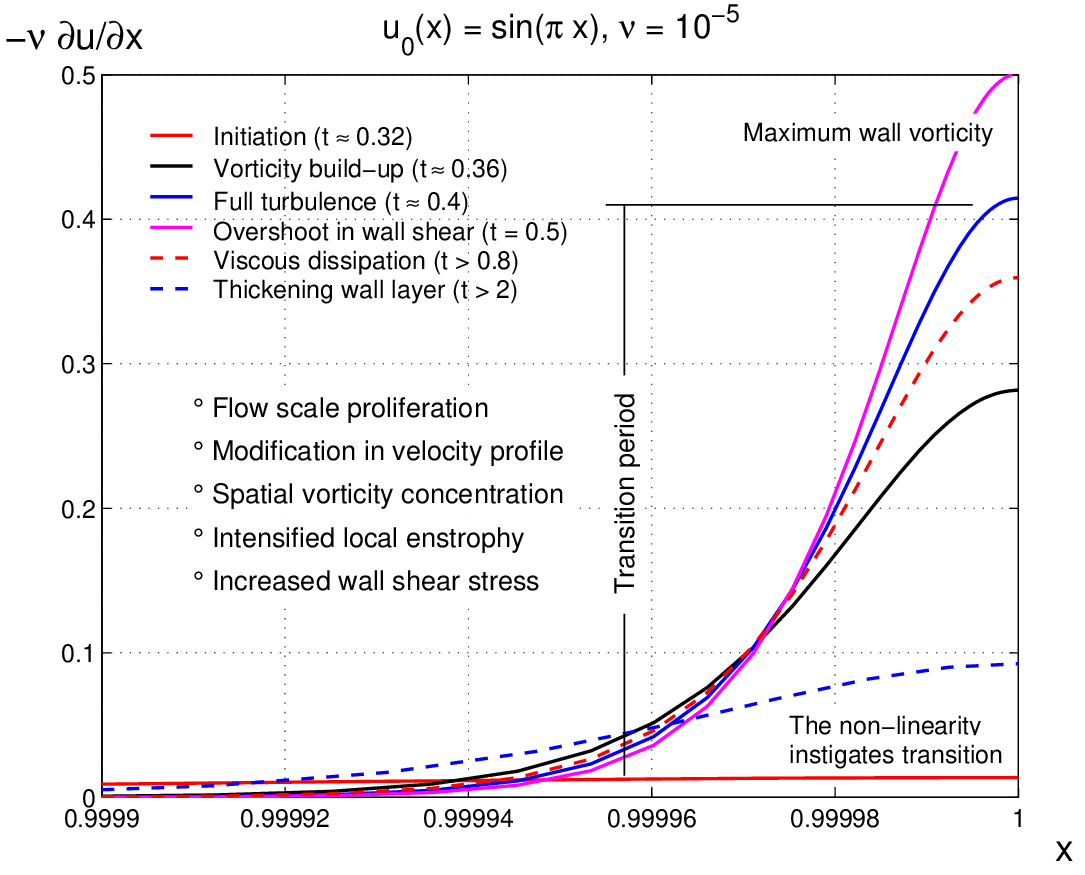}} 
 \caption{Refer to figure~\ref{bludu}. The intense build-up of vorticity at the wall between $t\approx 0.32$ and $t \approx 0.4$ is the hallmark of the laminar-turbulent transition. The continued shear growth from $t > 0.4$ to the maximum produces an overshoot in the local `skin friction' due to different rates of vorticity intensification across $x$. The state of the flow is fully turbulent as the friction is substantially higher than that before $t=0.32$. The fully-developed turbulence resides in the thin region $\Delta x \approx \nu$. In the subsequent decay, the viscous dissipation of the coalesced wall vorticity is largely confined to the wall vicinity $0.999 \leq x \leq 1$; the numerical data show that the maximum vorticity $|\partial_x u|_{\max} \approx 78, 20, 5$ and $0.2$ at time $t=25, 50, 100$ and $500$ respectively.  As $t \rightarrow \infty$, the viscous layer becomes thicker and expands toward the other end $x=0$, until the whole region $0 \leq x \leq 1$ reverts to a quiescent state. At $t=2000$, $u_{\max}$ decreases to $4.4{\times}10^{-4}$ near $x=0.9$, and the vorticity to $1.2{\times}10^{-2}$ at the wall. The overall shear is reduced to $5{\times}10^{-4}$ over $0 \leq x \leq 0.8$. } \label{rkduc} 
\end{figure}

\begin{figure}[ht] \centering
  {\includegraphics[keepaspectratio,height=10.5cm,width=15cm]{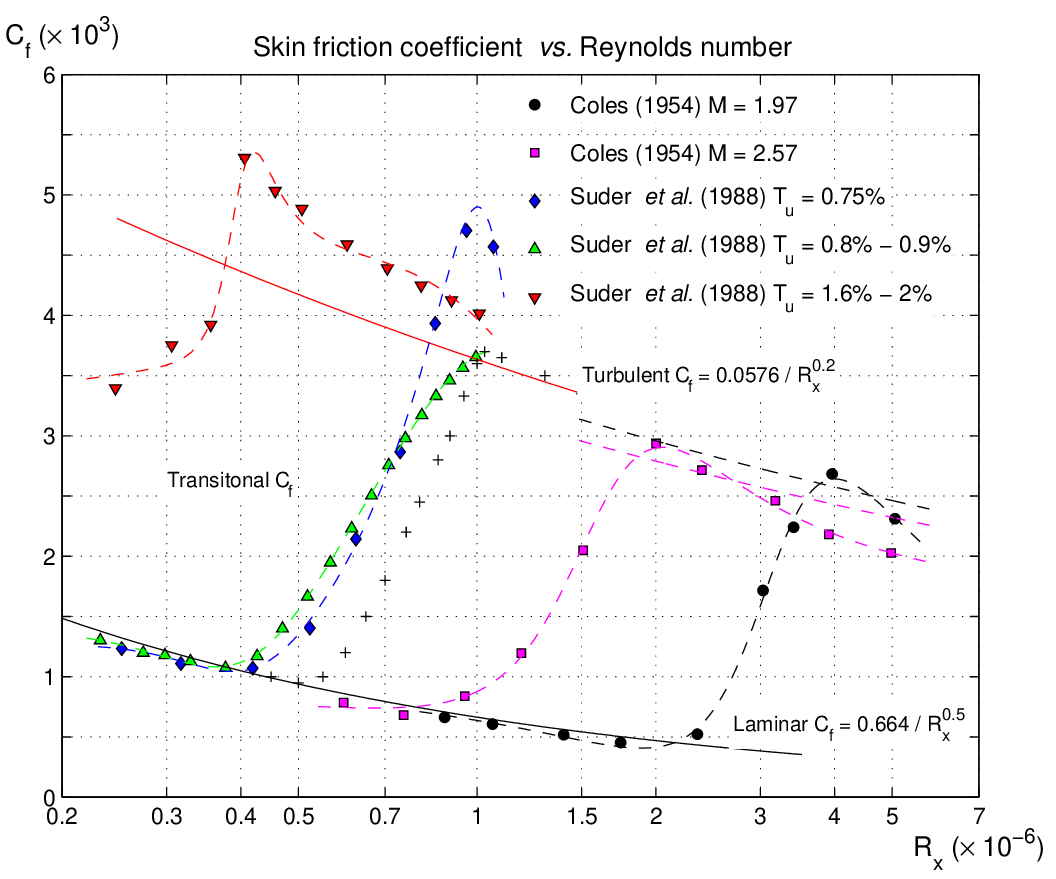}} 
 \caption{Increase in wall shear from laminar to turbulent state and evidence of overshoots in boundary layer experiments. Skin friction coefficient $C_f$ is defined as the ratio of wall shear $\tau_w$ and the free-stream dynamic head $\rho u^2_{\infty}/2$. Reynolds number $R_x$ is based on the stream-wise distance from the leading edge ($x$). The trend lines are added to the test data for indication only. Coles' data were collected over the natural transition. The M=1.97 data have been shifted by -0.1 for consistency. The grid-produced free-stream turbulence in the data of Suder {\it et al.} is measured in the fraction of average turbulent fluctuation relative to $u_{\infty}$ ($T_u$). This specific experiment deals with distorted transitions. Turbulent skin friction is calculated by assuming a velocity profile of $u/u_{\infty}=(y/\delta)^{\frac{1}{7}}$. The effects of Mach number on the skin friction are estimated and shown in broken curves. Particularly, the influence of the free-stream turbulence on the natural transition was far from clear. The last data set suggests that the life-span of the laminar boundary layer was rather short, if it ever existed. Nevertheless, each test run highlights a well-marked `ramp' over the transition, even the fully-developed turbulence downstream of the maximum stress behaves somehow differently. It is plausible that the incompressible data are not finely-tuned as the short-term records have to be dominated by the details in respective leading-edge condition. For reference, the trend (+ symbols) is added: it is the flat-plate natural transition starting $R_x\approx 0.5{\times}10^6$, estimated from the data in fig.II.25 of Lighthill (1963) and fig.21.2 of Schlichting (1979). The test results can be better understood by comparison with the calculations in figures~\ref{tevoamp} and \ref{tevocmp}.} \label{cfexpt} 
\end{figure}

It is established in laboratory experiment, that there is an abrupt growth in wall shear over transition region, see, for example, Schubauer \& Klebanoff (1955). At a free-stream turbulence level of $0.03\%$, the observed laminar-turbulent transition cannot be distorted by the free-stream turbulence intensity; the test transition is therefore the intrinsic transition embedded in the Navier-Stokes equations. From figure~3 of Schubauer \& Klebanoff, the shear increase occurs over a time interval of $3{\times}10^{-3}$ {\ttfamily s}, given a uniform free-stream speed of $24.38$ {\ttfamily m/s}. Moreover, referring to their figure~4, the intense vorticity concentrates in a spatial length scale of $1.5$ {\ttfamily mm}. Comparing the measurements (figure~\ref{skexpt}) and the numerical simulation (figure~\ref{rkduc}), the measured abrupt increase in wall shear is in qualitative agreement. The scale proliferation anticipates the time-wise occurrence of the shear build-up (the laminar-turbulent transition). The jump in friction over the transition zone (figure~\ref{cfexpt}) is a direct consequence of the non-linear growth.
\section{Discussion}
Even though Burgers' equation is merely one-dimensional, its solutions provide convincing evidence on the process of the scale proliferation which is essentially responsible for the abrupt local vorticity intensification.

Turbulence is characterised by vortices of multitudinous scales where its topologies encompass a variety of geometry, subject to continuous modifications by momentum transfer and diffusion. To describe turbulence in terms of Fourier components requires justification, because the flow scales defined by means of the wave components are certainly misguided portrayals of the genuine spatio-temporal structure. A critical pr\'ecis on this fundamental issue is given in \S 6.6.4 of Davidson (2004). 

On the continuum, the propensity for free deformation characterises fluids and is a precursor to resisting instability. Analytically, motions of both laminar and turbulent states are governed by the Navier-Stokes equations, {\it not} by the Ginzburg-Landau or Kuramoto-Sivashinsky equation. Solutions of the full equations are globally regular and hence they do not bifurcate in space and in time. Specifically, the solutions cannot develop any finite-time singularity for any initial data of bounded energy. It has been established in careful experiments that, the laminar-turbulent transition {\it persists} once the free-stream turbulence, or the acoustic noise, are below some threshold level (see, for intance, Schubauer \& Skramstad 1947; Wells 1967). The intensive local accumulation of vorticity is not connected to absolute instability engendered by spatially-amplified disturbances at selective frequencies; the apparent flow breakdown at fixed spatial locations observed in experiments is nothing more than a misreading of the fine-scale dynamic evolution.  

Although the intrinsic transition is fully instigated by the non-linearity, disturbances may be abundant in varied forms in practice. Frequently, an observed transition is an aberration of the non-linear process. However, the disturbances can be quantified and specified as part of the initial or boundary conditions; the modified problem is well-defined, and hence computationally tractable.\footnote{Our view contradicts those of instability theory, which claims success in a few incompressible cases. Taylor's paper on Couette flow between co-axial rotating cylinders (Taylor 1923) is not an instability analysis, but the {\it bona fide} method of eigen-function expansions applied to the linearised equations. The `disturbances' consist of non-amplifying Fourier components. There are no particular reasons to suspect its success over the initial phase of flow development in that specific topology. In the flat-plate experiment of Schubauer \& Skramstad (1947), artificial disturbances of known frequencies were intentionally introduced. The rationale for the good comparison with linear stability theory is that the Blasius mean flow is a credible {\it non-linear solution} in view of the boundary layer approximations (Prandtl 1904). Thus the infinitesimally perturbed laminar layer remains well-estimated by linearisation over the low-Reynolds-number regime close to plate's leading edge where the hypothesised local similarity holds. } 

In application, we may consider the effect of external control by adding a prescribed force, $f(x,t)$, to Burgers' equation. Similarly, the no-slip boundary conditions may be suitably modified in design of flow-control. Since the equation resembles the vorticity equation for two-dimensional incompressible viscous flows, investigation of the vorticity evolution at high Reynolds numbers looks promising. In fact, our numerical scheme needs minor modifications so as to embrace vorticity stretching $(\omega.\nabla) \bfu$ in three-dimensional space. As expected, the computation costs will be more substantial but would no longer be prohibitively expensive. The non-local nature of vorticity on velocity has to be taken into account, and its effect is expected to reduce the overall shear strength.
\section{Conclusion}
Burgers' equation is innocuous but is a well-chosen model for the full Navier-Stokes equations, as they share many of the essential analytic characters. From the present study, we are able to advance our understanding in one critical aspect: the governing equations furnish a reliable tool for predicting the laminar-turbulent transition, not only for the onset of the phenomenon but for the complete process. In an averaged sense, the transition is an intrinsic part of the flow evolution driven by the non-linearity and the energy conservation. In reality, viscous dissipation occurs at all times and stirs up randomness when the vorticity proliferation is amply operational. Turbulence is the incessant assembly of material subdivision. It is a fundamental fact that the transitional dynamics is Galilean invariant. 

Real shock waves and viscous wall layers have finite thickness $\sim O(\nu)$. Viscosity is the key, as it determines the spatial scales in the shear layers and irons out discontinuities. We summarise the development of shear flows in several generic stages:
(1) The initial shear growth is almost linear in time at all values of viscosity;
(2) There exists a time interval in which shear strength in localised region abruptly increases. The surge in vorticity due to the non-linearity marks the birth of turbulence;
(3) The end stage is easily identified when the local enstrophy attains a maximum value;
(4) The flow is fully turbulent and subsequently hibernates in a lengthy decaying period, particularly at small viscosity. The decay is governed by power-laws of time. 

The mathematical form of equation (\ref{beqn}) is identical to the KPZ equation (Kardar {\it et al.} 1986) for material interface growth or erosion due to vapour deposition. In astrophysics, suggestion has been put forward to use $3D$ Burgers' equations to model long-time self-organisation of the large structure in the Universe (see, for example, Woyczy\'nski 1998). The model is expected to be able to simulate formation of cellular structures in mass distribution.

No claims have ever been made on the validity of the continuum hypothesis at all spatio-temporal scales. It is well-established, from physics, that diffusion equation and the equations of general relativity have their limitations at atomic scales or are surely invalid at the Planck scale of length and time. Ironically, the global regularity of equation (\ref{beqn}) and of (\ref{ns}) asserts that these equations of fluid motion hold beyond the Planck scales for finite $\nu > 0$. The essence is that Burgers' equation or its Navier-Stokes progenitor offers a trustworthy model for the understanding of numerous practical problems formulated within the paradigm of Newtonian mechanics and classical thermodynamics. 
%
%
%
\addcontentsline{toc}{section}{\noindent{References}}

\begin{acknowledgements}
\vspace{10mm}

\noindent 
19 December 2018

\noindent 
\texttt{f.lam11@yahoo.com}
\end{acknowledgements}
%
%
%
%
%
\newpage
%
%
\label{lastpage}
\end{document}